\documentclass[letterpaper,twocolumn,10pt]{article}
\usepackage{authblk}

\pdfoutput=1

\newif\ifsubmit
\submittrue
\usepackage{soul}
\setstcolor{red}
\usepackage[table,svgnames]{xcolor}

\usepackage{silence}
\ifsubmit
\else
\fi

\usepackage{usenix-2020-09}
\usepackage{paralist}
\usepackage[scaled=.8]{FiraMono}

\newif\ifshownotes
\shownotesfalse
\newif\ifshowtodos
\showtodosfalse

\newcommand{\mytitle}{Systematic Assessment of Fuzzers using Mutation Analysis}

\usepackage{syntax}

\usepackage{hyperref}
\hypersetup{colorlinks=true,citecolor=blue,linkcolor=blue,filecolor=blue,urlcolor=blue,pdftitle={\mytitle}}

\definecolor{codegreen}{rgb}{0,0.6,0}
\definecolor{codegray}{rgb}{0.5,0.5,0.5}
\definecolor{codepurple}{rgb}{0.58,0,0.82}
\definecolor{backcolour}{rgb}{0.95,0.95,0.95}
\usepackage{listings} 
\usepackage{mdframed}

\definecolor{diffstart}{named}{Grey}
\definecolor{diffincl}{named}{Green}
\definecolor{diffrem}{named}{OrangeRed}

\usepackage{listings}
  \lstdefinelanguage{diff}{
    basicstyle=\ttfamily\small,
    morecomment=[f][\color{diffstart}]{@@},
    morecomment=[f][\color{diffincl}]{+\ },
    morecomment=[f][\color{diffrem}]{-\ },
  }

\usepackage[
    style=numeric,
    sorting=none,
    doi=false,
    isbn=false,
    url=false,
]{biblatex}
\renewbibmacro{in:}{}

\AtEveryBibitem{%
    \clearlist{location}
    \clearlist{organization}
    \clearfield{day}
    \clearfield{month}
    \clearfield{number}
    \clearfield{eventtitle}
    \clearfield{pages}
    \clearfield{publisher}
}

\renewcommand*{\bibfont}{\small}

\usepackage[most]{tcolorbox}

\usepackage{csquotes}
\usepackage{nicefrac}
\usepackage{siunitx}
\usepackage{array,framed}
\usepackage{booktabs}
\usepackage{
  color,
  float,
  epsfig, 
  wrapfig,
  graphics,
  graphicx,
  comment,
}
\usepackage{subcaption}
\usepackage{placeins}
\usepackage{framed}
\colorlet{shadecolor}{gray!40}

\usepackage{setspace}
\usepackage{latexsym,fancyhdr,url}
\usepackage{longtable}
\usepackage{algorithm2e}
\usepackage{algpseudocode}
\usepackage{graphics}
\usepackage{xparse} %
\usepackage{xspace}
\usepackage{multirow}
\usepackage{csvsimple}
\usepackage{balance}
\usepackage{tabularx}
\usepackage{xtab}
\usepackage{rotating}
\usepackage[inline]{enumitem}

\usepackage{
  tikz,
  pgfplots,
  pgfplotstable,
}

\usepackage{pifont}

\newcommand*{\circledRed}[1]{\textcolor{red}{\ding{\numexpr171 + #1}}}

\newenvironment{result}{\begin{framed}\centering\em}{\end{framed}}

\usepackage{hyperref}

\usetikzlibrary{
  shapes.geometric,
  arrows,
  external,
  pgfplots.groupplots,
  matrix,
  positioning
}

\pgfplotsset{compat=1.9}

\pgfplotsset{
  label style={font=\scriptsize},
  tick label style={font=\scriptsize},
  legend style={font=\scriptsize},
}

\definecolor{Butter}{HTML}{edd400}
\definecolor{Orange}{HTML}{f57900}
\definecolor{Chocolate}{HTML}{c17d11}
\definecolor{Chameleon}{HTML}{73d216}
\definecolor{SkyBlue}{HTML}{3465a4}
\definecolor{Plum}{HTML}{75507b}
\definecolor{ScarletRed}{HTML}{cc0000}
\definecolor{Aluminium}{HTML}{555753}

\makeatletter
\newcommand\resetfivestackedplots{
\makeatletter
\pgfplots@stacked@isfirstplottrue
\makeatother
\addplot [forget plot, draw=none] coordinates{
    (0,0) (1,0) (2,0) (3,0) (4, 0) 
};
}
\makeatother

\usepackage{mathtools,}

\DeclareMathAlphabet{\mathcal}{OMS}{cmsy}{m}{n}

\DeclareGraphicsExtensions{%
    .png,.PNG,%
    .pdf,.PDF,%
    .jpg,.mps,.jpeg,.jbig2,.jb2,.JPG,.JPEG,.JBIG2,.JB2}

\usepackage{acro}
\DeclareAcronym{asan}{
	short={ASAN},
	long={Address Sanitizer},
	cite={sere2012asan}
}
\DeclareAcronym{llvm-ir}{
	short={LLVM IR},
	long={LLVM Intermediate Representation},
	cite={llvm2022llvmlanguage}
}

\usepackage{cleveref}

\newcommand{\blackbox}{black-box\xspace}

\newcommand{\whitebox}{white-box\xspace}

\newcommand{\mua}{mutation analysis\xspace}
\newcommand{\Mua}{Mutation analysis\xspace}
\newcommand{\MuA}{Mutation Analysis\xspace}

\def\|#1|{\textit{#1}}
\def\<#1>{\texttt{#1}}
\def\[[#1\]]{\texttt{#1}} %

\definecolor{nontermcolor}{rgb}{0.4, 0.05, 0.0} %
\definecolor{absnontermcolor}{rgb}{0.4, 0.5, 0.0} %
\definecolor{evokcolor}{rgb}{0.8, 0.05, 0.1}    %
\definecolor{incorrectcolor}{rgb}{0.5, 0.0, 0.13}
\definecolor{incompletecolor}{rgb}{0.0, 0.0, 0.55}
\definecolor{validcolor}{rgb}{0.33, 0.42, 0.18} %
\definecolor{retcolor}{rgb}{0.65, 0.16, 0.16}
\definecolor{ipatterncolor}{rgb}{0.0, 0.5, 0.4} %
\definecolor{termcolor}{rgb}{0.0, 0.05, 0.4}    %
\definecolor{regexcolor}{rgb}{0.1, 0.3, 0.1}    %

\renewcommand{\ulitleft}{\normalfont\ttfamily}
\renewcommand{\litleft}{\bgroup\color{termcolor}`\ulitleft}
\renewcommand{\litright}{\ulitright'\egroup}

\renewcommand{\syntleft}{\bgroup\color{nontermcolor}$\langle$\normalfont\itshape}
\renewcommand{\syntright}{$\rangle$\egroup}

\newenvironment{densegrammar}{%
\begin{small}
\begin{grammar}%
}%
{%
\end{grammar}%
\end{small}%
}
{\begin{itemize}\item[]\begin{densegrammar}}%
{\end{densegrammar}\end{itemize}}%

\newcounter{todocounter}
\newcommand{\todo}[1]{\marginpar{$|$}\textcolor{red}{\stepcounter{todocounter}\footnote[\thetodocounter]{\textcolor{red}{\textbf{TODO }}\textit{#1}}}}

\newcommand{\rem}[1]{\textcolor{red}{\textbf{REMOVED }\st{#1}}}
\newcommand{\done}[1]{\marginpar{$*$}\textcolor{green}{\stepcounter{todocounter}\footnote[\thetodocounter]{\textcolor{black}{\textbf{DONE }}\textit{#1}}}}

\ifsubmit
\renewcommand{\todo}[1]{}
\renewcommand{\done}[1]{}
\renewcommand{\rem}[1]{}
\fi

\newcommand{\g}[1]{\cellcolor{yellow}#1}

\usepackage{outlines}

\newcounter{cenumi}
  \newcounter{cenumisaved}
  \newcommand{\labelcenumi}{}
    {\begin{list}{\labelcenumi}{\usecounter{cenumi}\partopsep=0pt\topsep=0pt\itemsep=0em\parsep=4pt\leftmargin=1em}%
    \setcounter{cenumi}{\value{cenumisaved}}}%
      {\setcounter{cenumisaved}{\value{cenumi}}%
    \end{list}}

\newcommand{\afl}{AFL\xspace}
\newcommand{\aflpp}{AFL++\xspace}
\newcommand{\libfuzzer}{libFuzzer\xspace}
\newcommand{\honggfuzz}{Honggfuzz\xspace}

\newcommand{\cares}{\texttt{cares}\xspace}
\newcommand{\caresname}{\texttt{cares\_name}\xspace}
\newcommand{\caresparsereply}{\texttt{cares\_parse\_reply}\xspace}
\newcommand{\vorbis}{\texttt{vorbis}\xspace}
\newcommand{\woffnew}{\texttt{woff2\_new}\xspace}
\newcommand{\libevent}{\texttt{libevent}\xspace}
\newcommand{\guetzli}{\texttt{guetzli}\xspace}
\newcommand{\retwo}{\texttt{re2}\xspace}
\newcommand{\curl}{\texttt{curl}\xspace}

\newcommand{\supermutant}{supermutant\xspace}

\newcommand{\supermutants}{supermutants\xspace}
\newcommand{\Supermutants}{Supermutants\xspace}

\begin{document}
\fancyhead{}
\title{\mytitle}

\author[1]{Philipp G\"orz}

\author[1]{Bj\"orn Mathis}

\author[1]{Keno Hassler}

\author[2]{Emre G\"uler}

\author[1]{\\Thorsten Holz}

\author[1]{Andreas Zeller}

\author[3]{Rahul Gopinath}

\affil[1]{CISPA Helmholtz Center for Information Security, Germany}
\affil[2]{Ruhr-Universität Bochum, Germany}
\affil[3]{University of Sydney, Australia}

\date{}

\maketitle

\begin{abstract}
Fuzzing is an important method to discover vulnerabilities in programs.
Despite considerable progress in this area in the past years, measuring and comparing the effectiveness of fuzzers is still an open research question.
In software testing, the gold standard for evaluating test quality is \emph{mutation analysis}, which evaluates a test's ability to detect synthetic bugs: If a set of tests fails to detect such mutations, it is expected to also fail to detect real bugs.
Mutation analysis subsumes various coverage measures and provides a large and diverse set of faults that can be arbitrarily hard to trigger and detect, thus preventing the problems of saturation and overfitting.
Unfortunately, the cost of traditional mutation analysis is exorbitant for fuzzing,
as mutations need independent evaluation.

In this paper, we apply modern mutation analysis techniques that pool multiple mutations and allow us---for the first time---to \emph{evaluate and compare fuzzers with mutation analysis}.
We introduce an \emph{evaluation bench} for fuzzers and apply it to a number of popular fuzzers and subjects. In a comprehensive evaluation, we show how we can use it to assess fuzzer performance and measure the impact of improved techniques. The required CPU time remains manageable: 4.09 CPU years are needed to analyze a fuzzer on seven subjects and a total of 141,278 mutations.
We find that today's fuzzers can detect only a small percentage of mutations, which should be seen as a challenge for future research---notably in improving
\begin{enumerate*}[label=(\arabic*), itemjoin={{; }}, itemjoin*={{ and }}, after={.}]
\item detecting failures beyond generic crashes
\item triggering mutations (and thus faults)
\end{enumerate*}
\end{abstract}

\section{Introduction}
\label{sec:intro}
Fuzzing is the key method to test the robustness of programs against malformed inputs.
Since it reveals inputs that crash or hang programs, and as these failures can often be turned into actual exploits, fuzzing is also the prime method to discover security vulnerabilities.
However, fuzzing is computationally expensive. %
Hence, researchers and practitioners must be able to determine which fuzzing tools and techniques are the most effective.
In a recent survey, 63\% of fuzzing practitioners~\cite{boehme2021fuzzing} named \emph{measures for fuzzer comparison} as one of the top three challenges that need to be solved.

Testing techniques, including fuzzers, are often assessed by obtained \emph{code coverage}~\cite{klees2018evaluating}.
Code coverage refers to the number of program elements that were exercised by the fuzzer (we treat code coverage in detail in \Cref{sec:background}).
This is reasonable because in order to find a bug at some location, the test must cover this very location in the first place.
But code coverage alone is not sufficient to actually find bugs.
Coverage on its own cannot evaluate the quality of sanitizers (used as bug oracles in fuzzing) or fuzzers set up to produce inputs specifically crafted to induce bugs~\cite{wang2010taintscope, osterlund2020parmesan}.
Furthermore, there is only a moderate association between bugs and coverage when using test generators: Previous research shows a correlation coefficient $R^2 = 0.72$ between Randoop generated test cases and mutation score~\cite{gopinath2014code}.
Another alternative to evaluate fuzzers is to run them on a benchmark of programs with \emph{known faults} and compare fuzzers by bugs found~\cite{klees2018evaluating,zhang2022fixreverter,hazimeh2020magma}.
A general concern with such approaches is that the distribution of available faults may not be uniform or related to the actual possible fault distribution in the program~\cite{bohme2022reliability}.
Furthermore, when the faults are known in advance, we run the risk of fuzzer parameters or even the technique itself being fine-tuned to find these faults~\cite{zeller2019when,zeller2019when2,bundt2021evaluating,boehme2021fuzzing}.

In software testing, the technique of \emph{mutation analysis} has established itself as the gold standard to evaluate tests and test generators~\cite{papadakis2019mutation}.
In mutation analysis, \emph{synthetic faults} (mutations) are injected into the program code by creating random variations (so-called \emph{mutants}).
The assumption is that a test set should be able to detect (``kill'') these mutations, just as it should be able to detect real faults.
As an example, \Cref{code:mutations} shows a number of possible mutations for a C code fragment.
We see that mutations such as changing the type of a variable (\circledRed{1}) or manipulating a comparison (\circledRed{2}, \circledRed{3}) may all impact the ability of a program to handle invalid inputs.
A good test set should be able to trigger these faults; and the more mutants a test set detects (``kills''), the higher its quality.

\begin{lstlisting}[label=code:mutations, belowskip=-1em, float=t, caption={
    Mutations in C code. Mutation \circledRed{1} deletes \texttt{unsigned}; mutation \circledRed{2} replaces \texttt{<} with \texttt{>=}; mutation \circledRed{3} adds \texttt{+16}.
    }]
<@\circledRed{1}~\st{\textbf{unsigned}}@> int len = message_length(msg);
if (len <@\circledRed{2}~\st{<}@> <@\textcolor{red}{>=}@> MAX_BUF_LEN <@\circledRed{3}~\textcolor{red}{+ 16}@>) {
    copy_message(msg);
} else {
    // Invalid length, handle error
}
\end{lstlisting}

\pagebreak
In contrast to coverage metrics, mutation analysis also assesses the ability of the tests to \emph{detect} the (injected) faults.
Indeed, a test can have 100\% coverage, but if it does not check any computation result, it will fail to detect errors.
In a fuzzing context, mutation analysis thus also assesses whether the fuzzer can detect issues beyond generic errors.
And in contrast to curated faults, tests cannot overfit, as the actual mutations being applied are many, diverse, and randomly distributed.
This lack of \emph{bias} in mutations (i.e., anything can happen, anywhere) is often touted as a big advantage of mutation analysis.
However, while it may be tempting to model mutants %
after past fixes~\cite{tufano2019icsme,zhang2022fixreverter}, this biases test assessment towards past issues, which in turn puts less weight on the ability of tests to find yet \emph{unknown} issues. (We are not aware how the Heartbleed~\cite{durumeric2014heartbleed}, ``goto fail;''~\cite{synopsis2014gotofail}, or log4shell~\cite{hunter2021log4shell} vulnerabilities could have been predicted from past issues.)

Several studies have confirmed the correlation between the ability to detect (mostly simple) mutations and the ability to detect (possibly complex) real faults~\cite{andrews2005icse,just2014fse,papadakis2018icse};
and several works have explored how mutation analysis can be applied to security testing~\cite{mouelhi2007mutation,dadeau2011mutationbased,woodraska2011security,le2007testing,loise2017towards,ami2021muse}.
Yet, mutation testing has a significant drawback:
It is \emph{expensive.}
Every single mutation induces a code change that needs to be evaluated by
an entire run of all tests to assess whether they detect the mutation---and this must be done for thousands of mutations.
Multiplied with the dynamic tests produced by fuzzers, this makes mutation analysis \emph{prohibitively expensive} for evaluating fuzzers.
Furthermore, fuzzers can react to faults~\cite{osterlund2020parmesan,wang2010taintscope,yu2022vulnerability}, limiting traditional avenues for optimizations in mutation analysis that assume static test suites.
Recent advances in mutation analysis, however, can significantly reduce this complexity.
Notably, the concept of a \emph{supermutant}~\cite{gopinath2018if} enables us to evaluate multiple mutants together in a single test run. The idea is to group together mutations that are unlikely
to interact and fuzz them in parallel.
Also, we can proceed in multiple phases, first having the fuzzer quickly achieve maximum coverage, producing a seed corpus for the next phase.
This seed corpus is then used as a static test suite to kill trivial mutants, leaving the few remaining (``stubborn'') ones as potential fuzzing targets.

In this paper, we show that such optimizations actually allow us to apply \emph{full mutation analysis to evaluate and compare fuzzers at scale,} using a myriad of \emph{unbiased mutations} that assess the ability of a fuzzer to find any kind of bugs, including, but not limited to, software vulnerabilities.
Our approach is implemented and available as a large-scale evaluation bench for fuzzers.
We demonstrate the usefulness of the evaluation bench by applying it on a number of popular fuzzers (\afl, \aflpp, \libfuzzer, and \honggfuzz) and test subjects
(\cares,
\vorbis,
\woffnew,
\libevent,
\guetzli,
\retwo, and
\curl).
We show that it is possible to conduct a full-scale evaluation and comparison of these fuzzers using 16.36 CPU years of computation time, just under a month with our available hardware---a nontrivial, yet affordable amount of resources and a first step toward\todo{towards? or am I wrong? -- BM} making mutation analysis a feasible solution for comparing fuzzers.
To the best of our knowledge, the present work thus
\begin{compactitem}
    \item is the first study to apply mutation analysis to fuzzers, using
    traditional mutation operators as well as operators from a security context
    (a total of 31 operators);
    \item identifies the limitations of traditional mutation analysis optimizations when used for
    fuzzer comparison;
    \item develops novel optimization strategies for mutation analysis when mutation analysis is
    used for fuzzer comparison (we show that these optimizations can make mutation analysis usable
    for fuzzer comparison);
    \item demonstrates that fuzzers indeed differ in their ability to detect mutations, and consequently, faults;
    \item shows that improvements in failure detection---notably the use of sanitizers---result in a better mutation detection.
\end{compactitem}
These results demonstrate that our evaluation bench can be used to compare fuzzer performance and evaluate improvements of different tools.

In our study, we also found that only a small percentage of mutants is detected by at least one of the fuzzers.
The best fuzzer in our evaluation, \aflpp, covers 30.9\% of all mutations and detects
28.3\% of these covered mutants---that is, 8.8\% of all mutations.
These numbers may seem low on an \emph{absolute} scale.
However, note that mutation analysis produces \emph{shallow mutants} that are easy to find, but also \emph{deep mutants} that are very hard to trigger, as well as \emph{subtle mutants} whose effects can be hard to detect.
There even are mutants that, although altering the \emph{syntax} of code or output, leave their \emph{semantics} untouched, and hence cannot be detected by any approach in the first place.
While a mutation detection rate of 100\% is thus unlikely to ever be achieved, the detection rate is very useful as a \emph{relative} measure to compare approaches and measure progress.
In our study, for instance, using an enhanced oracle such as \ac{asan} results in a moderate improvement in detection to 34.2\% of covered mutants (9.1\% of total mutants).

We find that our results reveal important directions for research in fuzzing:
\begin{compactitem}
    \item First, \emph{fuzzers can profit from better oracles}---that is, predicates that check for the presence of failures.
    Right now, fuzzers that use \emph{generic oracles}, such as crashes or hangs,
    are quite limited because not every vulnerability (and not every mutation)
    manifests itself this way.
    Our manual analysis shows that out of the mutants that were not detected,
    few could have been found by a crash oracle, see \Cref{sec:manual-analysis-mutations}. 
    However, a majority of these mutants would \emph{still produce a behavioral divergence}
    from the original program and hence can represent a vulnerability in a security context.
    \item Second, \emph{fuzzers can profit from more targeted intelligence}---that is, generating inputs based on possible bug locations.
    Out of all mutants killed by \aflpp (28.3\% of covered mutants),
    94.4\% were found with inputs generated on an unmutated binary.
    That is, at most 5.6\% of the mutants were killed because of targeted intelligence such as crafted inputs~\cite{wang2010taintscope} or directing
    fuzzers towards potential problems~\cite{osterlund2020parmesan,yu2022vulnerability}.
    This is significant because, according to an analysis of our syntactic fault patterns, we find that 71 of 100 recent~CVEs were \emph{coupled} \todo{Avian recommends to use ``related'' instead, but afaik, coupled is the technical term? -- KH} to a mutation.
    Hence, fuzzers could significantly benefit from better targeting.
\end{compactitem}

In summary, mutation analysis provides \emph{ambitious} and \emph{unbiased} goals for fuzzing and testing that are not susceptible to saturation or potential overfitting---and thus pose great challenges for future research.
Our optimizations of mutation analysis, as introduced in this paper, provide researchers and practitioners with the means to determine whether and which fuzzers meet these challenges.
Our evaluation bench is available at {\small\url{https://github.com/CISPA-SysSec/mua_fuzzer_bench}} to help assess future progress in the field.

\section{Technical Background}
\label{sec:background}
We use the IEEE~1044~\cite{ieee2010ieee} nomenclature:
A \emph{fault} is a code artifact causing a \emph{failure}.
A \emph{failure} (aka \emph{bug}) is an incorrect program behavior.
An \emph{error} is a human action that led to the fault.
An \emph{error model} defines the kinds of faults expected.

Given a limited computing budget, a fuzzing practitioner needs to choose a fuzzer
that is most likely to find the most bugs, which is typically accomplished
using \emph{coverage criteria} or a set of curated bugs~\cite{bohme2022reliability},
which we discuss next.

\subsection{Coverage Criteria for Fuzzer Comparison}
Coverage, or code coverage, refers to the number of program elements in the program under test (PUT) that were exercised by the input or set of inputs.
Some program elements typically considered are statements, basic blocks, branches, and unique acyclic paths through the program.
The idea is that code coverage provides an indication of the amount of program code that was explored.

Numerous coverage criteria exist~\cite{gopinath2014code} that can be used for
judging the effectiveness of test suites and test generators, such as fuzzers.
Most feedback-driven fuzzers, such as \afl, use some form of code coverage for guidance.
Hence, code coverage achieved in target programs can be seen as a
reasonable criterion for comparing fuzzers.
The main problem with using code coverage, however, is that it is insufficient on its own for evaluating fuzzers.
In particular, code coverage is unable to judge the quality of oracles
such as sanitizers.
They make bugs detectable (e.g., \ac{asan} detects many kinds of memory handling errors), which is critical to ensure effective fuzzing.
Another limitation is coverage saturation. 
That is, once a program element is saturated, there is little extra information available~\cite{chen2020revisiting}.
Many fuzzers~\cite{osterlund2020parmesan,mantovani2022fuzzing,wang2010taintscope}
include intelligence to craft inputs (e.g., calling an API with invalid values),
which, while important, is invisible if coverage is used for fuzzer comparison.

\subsection{Benchmarks Using Curated Faults}

Fuzzers can be evaluated using curated fault
benchmarks~\cite{klees2018evaluating,gavitt2016lava,hazimeh2020magma,metzman2021fuzzbench}.
However, such benchmarks are inherently limited to known faults.
As specific benchmarks are used to measure the effectiveness of a technique, the published improvements in the technique can become %
influenced by the benchmark in non-obvious ways. %
For example, if faults are mined from existing ones, the result might be numerous faults or types of faults that are easy to detect.
Further, if a given tool (such as \afl) was used to find and eliminate faults during development,
these faults are no longer in the set mined from released versions---%
which does not mean that the effectiveness of the tool has reduced.
For example, exchanging \afl for another tool that does well on such mined faults (but not necessarily on the faults already removed during development) may not produce the improvement a practitioner was hoping for.

That is, both bug-based and coverage-based techniques have inherent limitations.
Next, we discuss how \mua can overcome these.

\subsection{\MuA}
\label{sec:mua}

\Mua is a key technique for evaluating the fault-revealing power of test suites on a given program.
It is the premier method of test suite evaluation in both the industry~\cite{petrovic2018state,beller2021would,petrovic2018industrial} and the research community~\cite{papadakis2019mutation}.
It is a \whitebox technique that can be used to evaluate test suites when the source code of the program under test is available.

\begin{figure}[t!]
\begin{minipage}{\columnwidth}
\begin{tcolorbox}[boxrule=0pt,sharp corners, width=\columnwidth,enhanced,interior style={color={lightgray!30!white}}]
    We define the following mutation related terms that we use throughout this paper:
    \todo{Avian: It seems to me that this box should have some kind of title and should be referred to somewhere in the text. Otherwise, as a reader, I don't know when to read it. It somehow interrupts my reading flow.}
    \begin{description}[leftmargin=0cm]
    \itemsep=3pt
    \item[Mutation.] A small syntactic change that can be induced
      in the program. %
    \item[Mutation operator.] Transformation pattern that
      describes how mutations are induced in the program. A mutation operator,
      when applied to a matching location in the program, will produce a \emph{mutant}.
    \item[Mutant.] A new program that contains differences
      (mutations) from the original. A \emph{first order} mutant contains only a
      single mutation. A \emph{higher order} mutant contains multiple mutations.
      A \emph{supermutant} contains all possible mutations that can be applied at once.
    \item[Trivial mutants.] Mutants that can be killed
      without targeted intelligence.
      That is, any input whose execution covers their location will kill them.
    \item[Stubborn mutants.] Mutants that remain alive even after
      coverage reached their mutation locations.
    \item[Intelligent mutants.] Mutants killed by fuzzers on
      individual evaluation (i.e., they are not killed simply by covering their location).
    \item[Equivalent mutants.] An equivalent mutant is a mutant that, while different from
    the original program syntactically, has the same semantics.
    \item[Mutant kill matrix.] A mutant kill matrix (and similarly mutant coverage matrix) is a matrix with test cases as columns and mutants as rows. A mutant that is killed by a test case is assigned \<1> in the relevant cell (else \<0>)~\cite{ammann2014establishing}.
    \item[Minimal test suite.] A minimal test suite with respect to a given set of mutants and a given test suite containing simple test inputs is the smallest subset of test inputs that is required to kill the \emph{exact same mutants} as the given test suite~\cite{ammann2014establishing}.
    \item[Minimal mutant set.] A minimal mutant set with respect to a given set of mutants and a minimal test suite is the smallest subset of mutants that is required to maintain the minimal test suite~\cite{gopinath2016measuring}.
    \end{description}
\end{tcolorbox}
\end{minipage}
\end{figure}

For \mua, we start with the following \emph{error model}:
Any token in a program is a possible location for a fault to exist, and faults are likely caused during transcription of the concept in the developer's mind to the code artifact.
Further, we assume that the developer uses automatic tools such
as compilers, which removes some categories of faults.
This gives us a way to generate possible faults with minimal
human bias%
\footnote{
There remains an unavoidable human bias due to the selection of fault types.
However, generating instances of a fault type avoids bias as it is done exhaustively.
}%
:
Generate all instances of a fault type for each source code element that will get past the compiler.
However, many faults in the real world can be complex, containing multiple sub-faults. 
Modeling them with complex faults containing multiple sub-faults can lead to a combinatorial explosion, which can be avoided by depending on two well-studied
axioms---the \emph{finite neighborhood hypothesis} and the \emph{coupling effect}.
The \emph{finite neighborhood hypothesis} (also called \emph{competent-programmer hypothesis})
states that faults, if present in the program, are within a limited edit distance away from the correct formulation~\cite{gopinath2014mutations}.
The \emph{coupling effect} claims that simple faults are coupled to complex faults, such that tests capable of detecting failures due to simple faults will, with high probability, detect the failures due to complex faults.
Hence, the probability of fault masking is very low~\cite{offutt1992investigations}.
Both axioms are well researched, with well-founded theory~\cite{wah2000theoretical,wah2001theoretical,wah2003analysis,gopinath2017theory}, and confirmed in real-world software~\cite{just2014are,gopinath2017theory,petrovic2021does,chekam2017empirical}.
With these two axioms, we can limit the faults we need to test.
This allows us to focus on changes to the smallest program elements, such as tokens and statements, and still expect that the created mutations are representative of real bugs.

Given this error model, the idea of \mua is to collect possible fault patterns
(a single fault pattern is called a \emph{mutation operator}),
identify possible faults in the program (called \emph{mutations}),
generate corresponding faulty programs (called \emph{mutants})
each containing a single \emph{mutation},
and finally evaluate each \emph{mutant}
separately using each fuzzer and check whether the fuzzer is
able to detect the changed behavior of the mutant
(called \emph{killing the mutant}).

In summary, we can use the huge body of work on mutation analysis as an effective method to compare fuzzers by using the number of mutants killed by each fuzzer as the criterion.

\subsubsection{Computational Requirements}
\label{subsubsec:computation}
Cost is a major concern with \mua, as each mutant needs independent evaluation.
Furthermore, for fuzzing, we need to evaluate each produced input independently on each mutant.
We cannot tell if a mutant will be killed by an input without executing the mutant on that input. 
Indeed, we cannot even assume that the fuzzer will produce the same input on both the original and the mutant because the fuzzer may detect the mutation in the program and take steps to induce failure on a perceived fault.
That is, the number of program executions effectively increases quadratically with program size.

There are several traditional optimizations to make mutation analysis less costly.
However, these techniques assume static test suites, %
which makes them inapplicable to fuzzers.
For example, the most effective optimization is to find the statements in the program that are covered by the specific tests in the test suite, then only run tests against mutants they cover (or that lead to a state
infection~\cite{just2014efficient}).
This technique is inapplicable to fuzzers because fuzzers are non-deterministic, and the possibility of introspection on the source code for input generation might result in different inputs for the original and the mutant (violating the clean program assumption ~\cite{chekam2017empirical}).
The same problem affects usage of \emph{weak mutations}~\cite{howden1982weak}, split stream
execution~\cite{tokumoto2016muvm, gopinath2016topsy}, equivalence modulo states~\cite{wang2017faster},
and function memoization~\cite{ghanbari2021toward}.
Thus, these traditional optimization techniques do not work well for fuzzers.
We describe in \Cref{sec:approach} how we still achieve a significant reduction of computation time by using \supermutants~\cite{gopinath2018if}.

\subsubsection{Residual Defects}
\label{sec:residual}

One of the main reasons for using \mua is that it provides the best estimate for the number of \emph{residual defects}.
The residual defects are defects that remain in a program after testing is completed and all found defects have been fixed~\cite{gopinath2022mutation, ahmed2016can}.
Undetected faults that mirror a mutation (a single incorrect token) are accounted for with our approach, as all mutations are applied to a program exhaustively.
The larger and more complex types of faults are also subsumed by these mutations due to the \textit{coupling effect} hypothesis (see \Cref{sec:mua}).
That is, the number of mutants that remain undetected tracks the number of residual defects closely and can be considered a \emph{true ordinal measure}~\cite{gopinath2022mutation,tao2011introduction} of the number of residual defects in a program.

We also note that the residual defect density can be estimated from the number of mutants found using statistical estimation tools, such as \emph{population estimation}~\cite{pezze2008software}
and \emph{species richness} estimation as suggested by B\"ohme~\cite{bohme2019assurances}.

\subsubsection{Design of Mutation Operators}
Mutations are typically modeled on human errors such as exchanging a token in the
source code for another or forgetting to add a statement.
Some traditional mutation operators are carefully chosen so that a test suite capable of detecting
the resulting mutants also satisfies statement, branch~\cite{li2009experimental},
data-flow~\cite{frankl1997all,kakarla2011evaluation}, and various logic criteria~\cite{kaminski2013improving}.
That is, the test objective represented by a given criterion can be satisfied by the detection of a subset of mutants~\cite{papadakis2019mutation}.
In addition to the traditional operators, operators that reflect
known fault patterns in specific domains~\cite{papadakis2019mutation} are also chosen.

\subsubsection{Equivalent Mutants}
One of the problems with traditional \mua is equivalent mutants. These are mutants that
are semantically the same as the original program. For example, in the following fragment,
\begin{lstlisting}
if (cache.has(key)) return cache.get(key);
return compute(key);
\end{lstlisting}
removing the cache check need not induce a failure.
Previous studies show that 10\% to 23\% of generated mutants could be equivalent~\cite{jia2011analysis,yao2014study}, which can limit the usefulness of the mutation score (the absolute number of mutants killed).

We do not expect equivalent mutants to be a concern (beyond computational expenditure) for the following reasons:
(1) In fuzzer \emph{comparison}, only the \emph{relative mutant kills} matter.
(2) We manually analyzed a random sample of 100 stubborn mutants and found only 11 equivalent mutants.
This is also in line with the literature~\cite{jia2011analysis,yao2014study}.
Our analysis (in \Cref{sec:manual-analysis-mutations}) provides statistical confidence ($89 \pm 6\%$ CI at 99\% CL) that most of the generated mutants induced faults.

\section{Approach}
\label{sec:approach}
We now describe how to evaluate fuzzers using \mua with security-relevant mutations.
This requires answers to three questions: Which mutations to apply? How to detect if they are found? And, how to reduce the required computational effort?

\done{Avian: It is clear for the reader that you gave the background information in order to base the rest of your paper on it. I suggest to remove this part of the sentence and, instead, write two sentences that outline what you are going to describe in the following subsections.}

\subsection{Selecting Mutation Operators}
As fuzzers are mainly used to detect security issues,
we create mutations that focus on generating vulnerable code to compare different fuzzers.
We started with the traditional mutation operators
representing common programming errors as the baseline.
Next, we went through the list of Common Weakness Enumerations (CWEs) for
C~\cite{mitre2021weaknessc} and C++~\cite{mitre2021weaknesscpp},
creating mutations for interesting and feasible vulnerability types.
Finally, we investigated recent CVEs for C and C++ projects and added
unrepresented mutations modeling vulnerabilities that have been reported.

Our set of patterns can be grouped into seven types of mutations:
\emph{Compare patterns} (for example, increasing the right-hand side of a signed $<=$ comparison),
\emph{memory patterns} (modifying calls to allocation or de-allocation functions to trigger out-of-bounds, double free, or use-after-free issues),
\emph{control flow patterns} (deleting function calls or flipping branch conditions),
\emph{assignment patterns} (deleting variable assignments, changing comparisons into assignments, etc.),
\emph{library call patterns} (e.g., provoking a failure to test if return values are checked),
\emph{synchronization patterns} (removing locking operations or making atomics non-atomic),
and \emph{arithmetic patterns} (for example, turning a signed variable unsigned).
The full list of currently supported mutation operators
can be found in \Cref{tbl:mutations} in the Appendix.

\subsection{Detecting Mutations}
Another essential component is the detection of true mutant kills.
To make this process robust against the possibility of targeted manipulation,
it must be performed in a manner that makes it challenging for fuzzers to cheat.
For instance, if we classify a mutant as killed when the fuzzer reports a crash,
it would be trivial for a fuzzer to just report a crash in every run.
Similarly, we cannot base this decision on a version of the subject that
has been instrumented by a fuzzer,
as the fuzzer could itself introduce a crashing change during the instrumentation process.
Thus, to confirm any input that a fuzzer reports as crashing,
the input is rerun on the non-instrumented version of the original as well as the mutant
(see \Cref{sec:implementation} for a detailed description).
A mutation is killed if the original exits with a different exit code than the mutant does.
We require the original to exit without an error signal to avoid the case
that there is a crashing bug in the original.

\subsection{Reducing Computational Requirements}
\label{sec:reducing-computational-requirements}
We reduce the computational requirements by using two techniques: Split Analysis and Supermutants.

\paragraph{Split Analysis}
The analysis of fuzzer performance is split into two phases:
\begin{enumerate}[label=\Roman*.]
\item How much of the program is the fuzzer able to cover?
\item Does the fuzzer identify injected faults?
\end{enumerate}

For Phase~I, we are only interested in finding the maximum obtainable coverage,
and for that, we use the fuzzer under test on the original program
to generate a set of seed files that cover as much of the program as possible
(called \emph{coverage seeds} from now on).
Since no known faults are present, the incentive for the fuzzer is purely to cover the maximum amount of source code.
Next, we use the coverage seeds as a static test suite, where we can apply traditional \mua optimizations
and quickly remove any mutants that are killed by the coverage seed files.
This allows us to eliminate trivial mutants (those that only need coverage to crash), which
are a significant chunk of the total set of mutants, with limited computational overhead.

In Phase~II, we use the coverage seeds as the starting point to fuzz the remaining stubborn mutants.
This ensures that if the fuzzer contains ``intelligence'' to recognize and target the inserted fault,
it can use that intelligence to find and kill the mutation.
This method accounts for fuzzers that go beyond coverage
and use advanced code analysis to guide fuzzing.

\paragraph{\Supermutants.}
\label{sec:supermutants}
We use an approach based on \supermutants~\cite{gopinath2018if} to evaluate mutants with
fewer computational resources than traditional analysis.
The basic idea is to identify independent mutations and combine them into
\supermutants to allow a sound evaluation of multiple mutations without fault interactions
for the compute cost of a single mutant.

We identify two mutations as independent if no seed input covers both mutations during execution (for more context, see \Cref{sec:implementation}).
Depending on whether mutations were covered in Phase~I, we create supermutants as follows:
Covered mutations are combined into supermutants if they are mutually independent.
Non-covered mutations form supermutants by randomly choosing 100 mutations.
In both cases, a function contains at most one mutation 
(due to a technical limitation of our mutation engine).

If during Phase~II, we find that a supermutant cannot be killed, we mark all mutations in this supermutant as alive.
Otherwise, we identify the particular input or test case that killed it.
If more than one of the included mutants was covered, indicating a derivation from the initial identification\todo{assumption of independence -- BM},
we split up the supermutant and re-run the crashing input on the resulting mutants.
The fuzzing process is restarted for any surviving mutants,
which removes the possibility of multiple mutations interfering with each other.

\section{Implementation}
\label{sec:implementation}

\begin{figure*}[ht]
  \begin{subfigure}{0.43\textwidth}
    \centering
    \includegraphics[width=\textwidth]{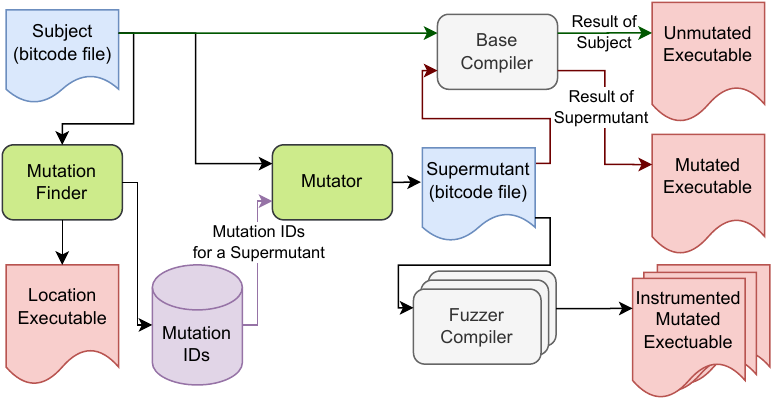}
    \caption{The different executables created from a single subject.}
    \label{fig:compilation}
  \end{subfigure}
  \hfill
  \begin{subfigure}{0.54\textwidth}
    \centering
    \includegraphics[width=\textwidth]{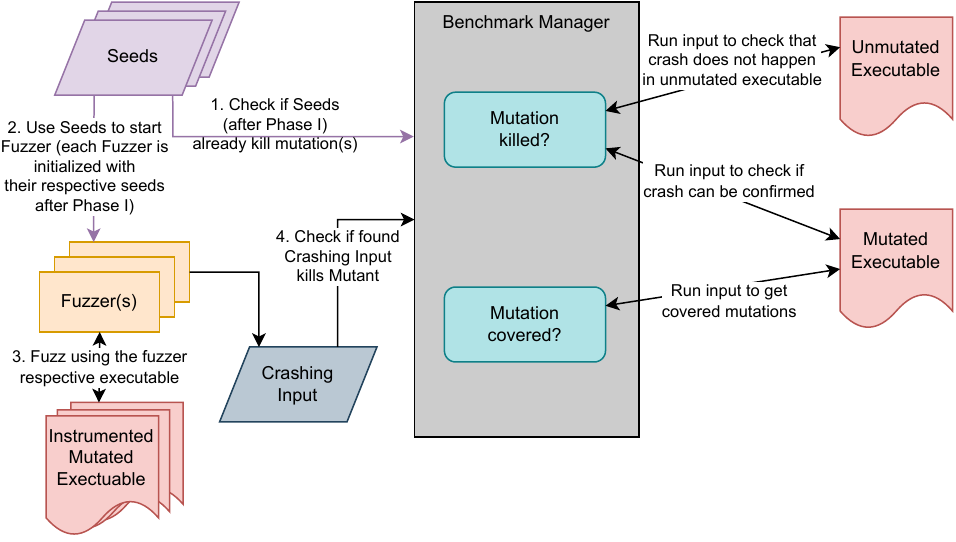}
    \caption{Steps taken to evaluate mutations.}
    \label{fig:evaluatingprocess}
  \end{subfigure}
  \caption{Schematic overview of the compilation and evaluation process.}
  \label{fig:schematic-overview}
\end{figure*}

Our chosen subjects are C and C++ projects to cover programs that are affected by memory corruption vulnerabilities. %
We implemented an LLVM pass to find mutation locations (\textit{mutation finder})
and another pass to do the actual code changes (\textit{mutator}).
See \Cref{fig:compilation} for an overview of the compilation process.
The mutation finder identifies all mutation locations and possible mutations
and assigns an ID to each of them.
It also produces an executable (location executable) with logging code in place of the actual mutation
that can be used to check which mutations are covered for a given input.
One comparison executable without any mutations is compiled using the base compiler.
Given mutation ID(s), the mutator produces the corresponding supermutant bitcode file.
The bitcode file is then compiled with a base compiler to create a mutated executable for comparison,
and each fuzzer compiles its own instrumented version.
To decide which mutations can be put together into one supermutant,
we run all seed inputs on the executable produced by the mutation finder.

To evaluate a mutation, we need to know if the mutation has been covered and whether it has been killed.
An overview of this process is shown in \Cref{fig:evaluatingprocess}.
During fuzzing, we check if a mutation is covered using the mutated executables.
To decide if a mutation has been killed, we rerun inputs that a fuzzer reports as crashing on the unmutated and mutated executables. %

\section{Evaluation}
\label{sec:eval}
Our evaluation seeks to answer the central question:
\emph{How well can fuzzers be compared using \MuA?}
We evaluate four research questions to study
mutation analysis as a comparative metric for fuzzer evaluation.

\begin{description}%
\item[RQ1.] \emph{How do different fuzzers compare in killing mutants?}
The question has three parts:
(1) what percentage of mutants were killed in Phase~I,
(2) what percentage of the remaining mutants were killed in Phase~II, and
(3) how do the killed mutant sets intersect between different fuzzers?

\item[RQ2.] \emph{How much can sanitizers improve the results?}
Inputs generated by a fuzzer account for only a part of the toolchain.
Detecting bugs requires some kind of oracle, ranging from simple crash feedback
to more sophisticated sanitizers.
With this question we want to assess if we can measure sanitizer influence.

\item[RQ3.] \emph{How many real vulnerabilities are coupled to mutations?}
For \mua to be useful, the mutations it produces should be semantically coupled to some real faults.
That is, for any real fault, there should exist a mutation such that detecting the mutation guarantees
detecting the real fault~\cite{just2014are}.
Hence, this question seeks to understand whether we have succeeded in capturing
the characteristics of real-world vulnerabilities with our mutation operators.

\end{description}

\subsection{Setup}
\label{sec:setup}
We describe our experimental setup next.

\paragraph{Subjects.}
For our experiments, we looked for robust subjects with no known faults.
We selected OSS-Fuzz~\cite{google2021ossfuzz} subjects (see \Cref{table:subjects}) that were compilable with gllvm and do not crash when fuzzed for 48 hours.

\begin{table}[tb]
\centering
\caption{Overview of the subjects used for evaluation. LOC as counted by cloc~\cite{aldanial2023cloc}, taking the sum of C/C++ lines.}
\small
\begingroup
\resizebox{\columnwidth}{!}{
    \begin{tabular}{@{} l l c r @{}}
    Subject           & Description          & Version  & LOC \\
    \midrule 
    \caresname        & DNS query creation   & 809d5e84 &     62,749 \\       
    \caresparsereply  & DNS reply parsing    & 809d5e84 &     62,749 \\
    \woffnew          & Font Format          & 4721483a &     39,373 \\
    \libevent         & Event Notification   & 5df3037d &     56,881 \\
    \guetzli          & JPEG Encoder         & 214f2bb4 &      8,029 \\
    \retwo            & Regular Expressions  & 58141dc9 &     27,545 \\
    \curl             & Data Transfer        & curl-7_83_1 & 151,413 \\
    \end{tabular}
}
\endgroup
\label{table:subjects}
\end{table}
\paragraph{Fuzzers.}
We chose fuzzers that are general purpose, support in-process fuzzing,
offer a compiler-wrapper and \ac{asan} support, resulting in
\afl, \aflpp, \libfuzzer, and \honggfuzz for evaluation (see \Cref{table:fuzzers}).
We follow generally recommended flags and use forking mode for \libfuzzer to avoid early stopping.
If dictionaries are available for subjects, they are provided to the fuzzers.

\begin{table}[tb]
  \centering
  \caption{
    Overview of the fuzzers used for evaluation. The arguments under Dictionary are additionally provided if there is a dictionary available.
  }
  \begin{tabular}{@{} l l l l @{}}
    Fuzzer     & Arguments           & Dictionary    & Version          \\
    \midrule
    \afl       & \verb|-d|           & \verb|-x|     & \phantom{0}2.57b \\
    \aflpp     & \verb|-d -c cmplog| & \verb|-x|     & \phantom{0}3.14c \\
    \honggfuzz & \verb|-n 1|         & \verb|--dict| & \phantom{0}2.4   \\
    \libfuzzer & \verb|-fork=1|      & \verb|-dict|  & 11.1             \\
  \end{tabular}
  \label{table:fuzzers}
\end{table}

\paragraph{Hardware.}
We used four servers providing an Intel Xeon Gold 6230R CPU, each with 52 cores and 188 GB RAM.
The computation times for all experiments are shown in \Cref{table:actual-compute-time}.

\begin{table}[tb]
  \centering
  \caption{
    Actual computation times for all experiments.
    Experiments were run on four servers, each with 52 cores.
    Note that evaluating a single fuzzer takes ~4.09 CPU years (\enquote{Seed + Default} / \#Fuzzers) with our chosen subjects.
  }
  \label{table:actual-compute-time}
  \begin{tabular}{ @{} l r r @{} }
                    & CPU (Years) & 4 Servers (Days) \\ \midrule
Seed Collection & 1.99        & 3.50            \\
Default         & 14.37       & 25.22           \\
Seed + Default  & 16.36       & 28.72           \\
ASAN            & 15.16       & 26.61           \\
24 Hours Runs   & 7.42        & 13.02           \\ \midrule
Sum             & 38.95       & 68.34           \\

  \end{tabular}
  \vspace{-1em}
\end{table}

\paragraph{Runtime.}
As the required CPU time is a central concern, we employ two approaches to reduce the required CPU time,
which are described in \Cref{sec:reducing-computational-requirements}.
One is the application of supermutants which, for our evaluation, results in a reduction of required runs by between $1.76\times$ and $19.95\times$ (a mean reduction of $3.8\times$), as shown in \Cref{table:compute-reduction}.
Note that this ignores supermutants that are not independent---see \Cref{table:independent} for details.

The other approach is splitting the benchmark into a coverage phase (I) and a fault-detection phase (II).
A speedup is achieved by using a long individual runtime for phase I (for our evaluation, 13 repetitions of 48 hours each) to allow a shorter runtime for phase II (1 hour per supermutant).
The real speedup is highly dependent on the chosen runtimes and can be calculated by the following formula:
\(\frac{M \times R}{R_I \times S + M \times R_{II}}\);
where M is the number of supermutants, $R$ is the normal runtime, $R_I$ is the runtime of phase I, $R_{II}$ is the runtime of phase II, and $S$ is the number of repetitions for coverage seed gathering.
For our evaluation, this results in a speedup of $25\times$.
\todo{In this section, we don't capitalize \emph{phase}. Why? -- KH}

\begin{table}[tb]
  \centering
  \caption{Computational Reduction by Using Supermutants
    \todo{KH: I modified the table to address Avian's comments, but this is not yet updated in R.}
  }
  \begin{tabular}{@{} l r @{ \ } r @{ \ } r @{}}
Subject          & \#Mutants & \#Supermutants & Factor \\ \midrule
\curl            & 29,118  & 5,804        & 5.02      \\
\guetzli         & 22,961  & 13,040       & 1.76      \\
\woffnew         & 40,914  & 5,930        & 6.90      \\
\caresname       & 4,822   & 550          & 8.77      \\
\caresparsereply & 4,822   & 1,288        & 3.74      \\
\libevent        & 17,234  & 864          & 19.95     \\
\retwo           & 21,407  & 9,670        & 2.21      \\
\midrule
Sum              & 141,278 & 37,146       & \textbf{3.80} \\

  \end{tabular}
  \label{table:compute-reduction}
  \vspace{-1em}
\end{table}

\subsection{RQ1: How do different fuzzers compare?}
\begin{table*}[!t]
\centering
\caption{
Results of the full benchmark run.
The \emph{\#Mutations} column contains the number of mutations available.
\emph{Phase~I} represents 24-hour runs,
\emph{Phase~II} represents the one-hour runs, and \emph{Total} represents both combined.
\emph{Covered} represents the number of covered mutants, while \emph{Killed}
represents the number of killed mutants.
The \textbf{combined} rows show the total number of mutants that were killed by
\emph{any} fuzzer.
}
\begingroup
   \fontsize{7pt}{4pt}\selectfont
    \begin{tabular}{ @{} l r l r r r r r r r r @{} }
    \multicolumn{2}{c}{} & \hspace{3mm} &  \\
    Program                            & \#Mutations            & Fuzzer                & Phase~I Covered & Phase~I Killed & Phase~II Covered & Phase~II Killed & Total Covered & Total Killed \\
\cmidrule{1-9} \\
\multirow{5}{*}{cares_name}        & \multirow{5}{*}{4822}  & afl                   & 88              & 17             & 0                & 3               & 88            & 20           \\
                                   &                        & aflpp                 & 88              & 18             & 0                & 2               & 88            & 20           \\
                                   &                        & honggfuzz             & 88              & 17             & 0                & 1               & 88            & 18           \\
                                   &                        & libfuzzer             & 86              & 17             & 0                & 0               & 86            & 17           \\
                                   &                        & \g{\textbf{combined}} & \g{88}          & \g{18}         & \g{{ }}          & \g{{ }}         & \g{88}        & \g{20}       \\ \cmidrule{3-9}
\multirow{5}{*}{cares_parse_reply} & \multirow{5}{*}{4822}  & afl                   & 937             & 292            & 0                & 29              & 937           & 321          \\
                                   &                        & aflpp                 & 941             & 289            & 0                & 26              & 941           & 315          \\
                                   &                        & honggfuzz             & 940             & 290            & 1                & 23              & 941           & 313          \\
                                   &                        & libfuzzer             & 932             & 292            & 0                & 5               & 932           & 297          \\
                                   &                        & \g{\textbf{combined}} & \g{941}         & \g{305}        & \g{{ }}          & \g{{ }}         & \g{941}       & \g{324}      \\ \cmidrule{3-9}
\multirow{5}{*}{curl}              & \multirow{5}{*}{29118} & afl                   & 9,935           & 2,328          & 89               & 68              & 10,024        & 2,396        \\
                                   &                        & aflpp                 & 11,713          & 2,593          & 69               & 82              & 11,782        & 2,675        \\
                                   &                        & honggfuzz             & 10,195          & 2,299          & 506              & 150             & 10,701        & 2,449        \\
                                   &                        & libfuzzer             & 8,895           & 2,099          & 120              & 39              & 9,015         & 2,138        \\
                                   &                        & \g{\textbf{combined}} & \g{12,459}      & \g{2,807}      & \g{{ }}          & \g{{ }}         & \g{12,477}    & \g{2,857}    \\ \cmidrule{3-9}
\multirow{5}{*}{guetzli}           & \multirow{5}{*}{22961} & afl                   & 6,943           & 1,691          & 6,189            & 1,827           & 13,132        & 3,518        \\
                                   &                        & aflpp                 & 12,564          & 3,205          & 685              & 378             & 13,249        & 3,583        \\
                                   &                        & honggfuzz             & 13,586          & 3,698          & 0                & 72              & 13,586        & 3,770        \\
                                   &                        & libfuzzer             & 9,923           & 2,816          & 32               & 26              & 9,955         & 2,842        \\
                                   &                        & \g{\textbf{combined}} & \g{13,610}      & \g{3,810}      & \g{{ }}          & \g{{ }}         & \g{13,610}    & \g{3,912}    \\ \cmidrule{3-9}
\multirow{5}{*}{libevent}          & \multirow{5}{*}{17234} & afl                   & 425             & 75             & 0                & 6               & 425           & 81           \\
                                   &                        & aflpp                 & 427             & 78             & 0                & 3               & 427           & 81           \\
                                   &                        & honggfuzz             & 421             & 77             & 1                & 3               & 422           & 80           \\
                                   &                        & libfuzzer             & 417             & 77             & 0                & 2               & 417           & 79           \\
                                   &                        & \g{\textbf{combined}} & \g{427}         & \g{80}         & \g{{ }}          & \g{{ }}         & \g{427}       & \g{81}       \\ \cmidrule{3-9}
\multirow{5}{*}{re2}               & \multirow{5}{*}{21407} & afl                   & 4,825           & 4,461          & 0                & 1               & 4,825         & 4,462        \\
                                   &                        & aflpp                 & 11,563          & 4,457          & 65               & 134             & 11,628        & 4,591        \\
                                   &                        & honggfuzz             & 11,636          & 4,431          & 0                & 97              & 11,636        & 4,528        \\
                                   &                        & libfuzzer             & 11,164          & 4,328          & 0                & 20              & 11,164        & 4,348        \\
                                   &                        & \g{\textbf{combined}} & \g{11,639}      & \g{4,571}      & \g{{ }}          & \g{{ }}         & \g{11,639}    & \g{4,627}    \\ \cmidrule{3-9}
\multirow{5}{*}{woff2_new}         & \multirow{5}{*}{40914} & afl                   & 5,386           & 955            & 2                & 70              & 5,388         & 1,025        \\
                                   &                        & aflpp                 & 5,578           & 1,044          & 1                & 69              & 5,579         & 1,113        \\
                                   &                        & honggfuzz             & 5,555           & 1,018          & 1                & 111             & 5,556         & 1,129        \\
                                   &                        & libfuzzer             & 5,018           & 896            & 2                & 20              & 5,020         & 916          \\
                                   &                        & \g{\textbf{combined}} & \g{5,631}       & \g{1,139}      & \g{{ }}          & \g{{ }}         & \g{5,634}     & \g{1,232}    \\

    \end{tabular}
\endgroup

\label{table:big-table}
\end{table*}

Seed inputs are first created for each fuzzer and subject pair.
These are either extracted from subject repositories if available or created manually.
Each fuzzer is run on the unmutated base executable of the subject to fuzz for coverage
for 48 hours (13 instances per fuzzer).
The resulting inputs are then minimized by the tools provided by the respective fuzzer.
Of the 13 instances for each fuzzer, the median run based on
covered mutations is selected as the coverage seed corpus. %
The median run is used to avoid outliers in performance,
as is recommended by previous research~\cite{klees2018evaluating}.

Next, we create \supermutants (\Cref{sec:supermutants})
and evaluate each on the coverage seeds from Phase~I.
In Phase~II, each fuzzer is run for an hour on each supermutant.%

\paragraph{Results.}

The results of this experiment are provided in \Cref{table:big-table}.
The column \emph{\#Mutations} represents the number of produced mutations.
\emph{Fuzzer} represents the used fuzzer.
\emph{Phase~I Covered} represents the number of mutations covered by the seeds.
\emph{Phase~I Killed} represents the number of mutations killed by the seeds.
\emph{Phase~II Covered} represents the number of mutations that were covered in the
Phase~II beyond Phase~I.
\emph{Phase~II Killed} represents the number of mutations that were killed in Phase~II beyond Phase~I.
\emph{Total Covered} represents the number of mutants that were covered in total,
and \emph{Total Killed} represents the number of mutants that were killed in total.
\done{KH: I removed the two previous sentences, since the same information is given again in the paragraph ``How do these fuzzers relate to each other?''}
Summing up all subjects, we see that \aflpp kills 8.8\% of all mutants,
closely followed by \honggfuzz with 8.7\% and \afl with 8.4\%.
\libfuzzer lags behind at only 7.5\%.
In terms of covered mutations, \aflpp (30.9\%) is again narrowly in front of \honggfuzz
(30.4\%), followed after a large gap by \libfuzzer (25.9\%) and \afl (24.6\%).
We find that coverage alone accounts for most of the killed mutants.
The ensemble of all fuzzers in our evaluation (\emph{combined} rows in
\Cref{table:big-table}) gets 99.95\% of its coverage and 97.5\% of its kills in Phase~I.
Indeed, none of the evaluated fuzzers employ bug-targeted feedback instrumentation, hence,
this is expected.
\begin{result}
Coverage accounts for most mutants (97.5\%) detected in our evaluation.
\end{result}
We found two anomalous results regarding \afl:
For \guetzli, the median run covers only around 7,000 mutations,
which is caused by wildly inconsistent runs covering from 2,500 to 12,000 mutations,
a disadvantage from which \afl recovers surprisingly well in Phase~II.
Additionally, for \retwo, \afl crashes for most mutations.
\done{KH: I removed the following sentence. It probably just confuses the reader, and also is just a guess.}

\begin{figure}[ht]
   \centering
   \includegraphics[width=\columnwidth]{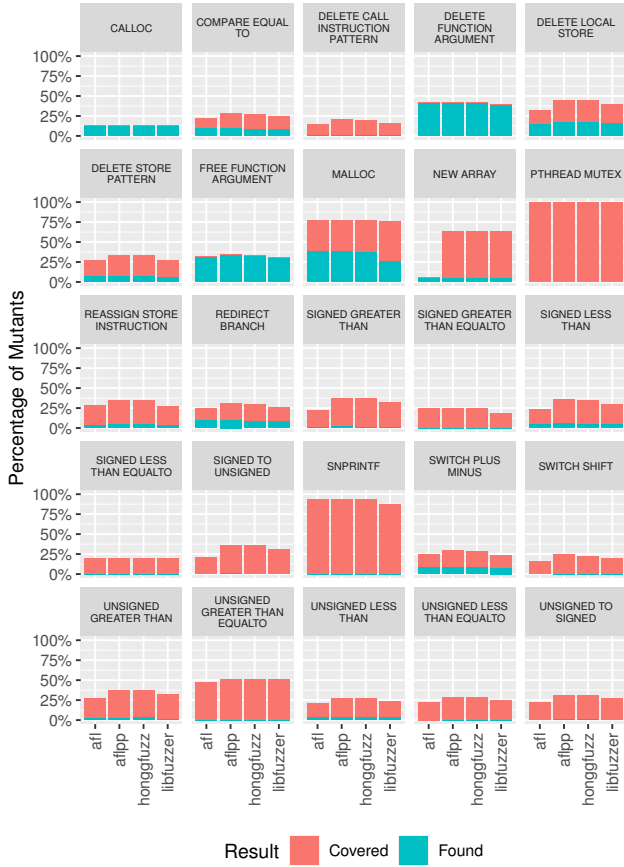}
   \caption{
    Percentage of covered and found mutants per mutation type and fuzzer (all found mutations are inherently covered).
    \afl underperforms for some mutation types due to crashes caused by instrumentation bugs. This can also be seen in \Cref{table:big-table} and \Cref{table:big-table-asan}.
   }
   \label{fig:mutation-types}
\end{figure}

\begin{figure}[tb]
    \centering
    \includegraphics[width=\columnwidth]{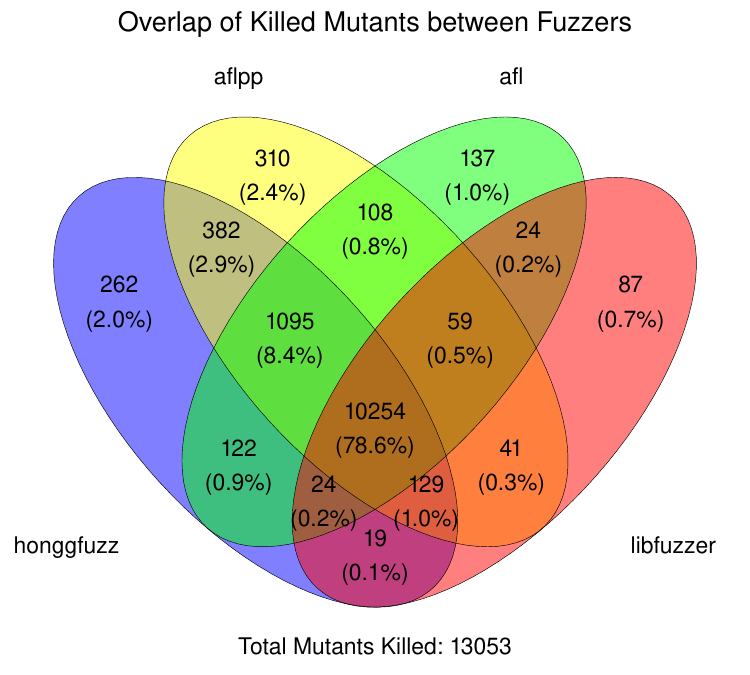}
    \caption{Venn diagram of killed mutants for each fuzzer.}
    \label{fig:wayne}
\end{figure}

\paragraph{How do these fuzzers relate to each other?}
See \Cref{fig:wayne} for a Venn diagram of killed mutants per fuzzer.
The fuzzers have a large intersection of killed mutations ($78.6\%$).
Consider \aflpp: It finds 94.9\% of all killed mutants.
Adding a second fuzzer provides only a marginal improvement: \honggfuzz (+3.2\%), \afl (+2.3\%) or \libfuzzer (+1.3\%).
For further details, a comparison per mutation type is available in \Cref{fig:mutation-types}.

\paragraph{Is an extra hour of fuzzing beyond 24 hours of fuzzing for coverage sufficient
for targeted fuzzing?}
To examine this question, we sampled up to 104 (to keep a multiple of CPU count)
stubborn mutants for each subject and fuzzed them for 24 hours on every fuzzer.
Note that, as we do not use \supermutants for this experiment (i.e., we use only one mutation per mutant),
this also serves as a test if \supermutants introduce inconsistencies.
As only \textit{three} of the total of 690 mutants are killed,
we believe that one hour runs are indeed sufficient for Phase~II.
Additionally, we feel that this result confirms our choice of using supermutants. 
The detailed results can be found in \Cref{table:24-hour-runs}.

\begin{table*}[!t]
\centering
\caption{Results of the benchmark run \emph{when ASAN is enabled}.
The \emph{\#Mutations} column contains the number of mutations available.
\emph{Phase~I} represents 24-hour seed gathering,
\emph{Phase~II} represents the one-hour runs, and \emph{Total} represents both combined.
The \emph{Covered} represents the number of mutants covered, while \emph{Killed}
represents the number of mutants killed.
The \textbf{combined} rows show the total number of mutants killed by
\emph{any} fuzzer.
}
\begingroup
   \fontsize{7pt}{4pt}\selectfont
    \begin{tabular}{ @{} l r l r r r r r r r r r @{} }
    \multicolumn{2}{c}{} & \hspace{3mm} &  \\
    Program                            & \#Mutations            & Fuzzer                & Phase~I Covered & Phase~I Killed & Phase~II Covered & Phase~II Killed & Total Covered & Total Killed \\
\cmidrule{1-9} \\
\multirow{5}{*}{cares_name}        & \multirow{5}{*}{4822}  & afl                   & 88              & 21             & 0                & 0               & 88            & 21           \\
                                   &                        & aflpp                 & 88              & 22             & 0                & 0               & 88            & 22           \\
                                   &                        & honggfuzz             & 88              & 21             & 0                & 1               & 88            & 22           \\
                                   &                        & libfuzzer             & 87              & 21             & 0                & 1               & 87            & 22           \\
                                   &                        & \g{\textbf{combined}} & \g{88}          & \g{22}         & \g{{ }}          & \g{{ }}         & \g{88}        & \g{22}       \\ \cmidrule{3-9}
\multirow{5}{*}{cares_parse_reply} & \multirow{5}{*}{4822}  & afl                   & 938             & 429            & 0                & 1               & 938           & 430          \\
                                   &                        & aflpp                 & 941             & 418            & 0                & 17              & 941           & 435          \\
                                   &                        & honggfuzz             & 940             & 427            & 1                & 7               & 941           & 434          \\
                                   &                        & libfuzzer             & 906             & 427            & 0                & 2               & 906           & 429          \\
                                   &                        & \g{\textbf{combined}} & \g{941}         & \g{432}        & \g{{ }}          & \g{{ }}         & \g{941}       & \g{438}      \\ \cmidrule{3-9}
\multirow{5}{*}{curl}              & \multirow{5}{*}{28779} & afl                   & 9,981           & 3,068          & 87               & 38              & 10,068        & 3,106        \\
                                   &                        & aflpp                 & 11,680          & 3,458          & 68               & 73              & 11,748        & 3,531        \\
                                   &                        & honggfuzz             & 10,155          & 3,057          & 481              & 162             & 10,636        & 3,219        \\
                                   &                        & libfuzzer             & 8,833           & 2,728          & 130              & 80              & 8,963         & 2,808        \\
                                   &                        & \g{\textbf{combined}} & \g{12,349}      & \g{3,729}      & \g{{ }}          & \g{{ }}         & \g{12,387}    & \g{3,774}    \\ \cmidrule{3-9}
\multirow{5}{*}{guetzli}           & \multirow{5}{*}{22961} & afl                   & 3,713           & 1,019          & 3,187            & 903             & 6,900         & 1,922        \\
                                   &                        & aflpp                 & 6,704           & 1,903          & 340              & 212             & 7,044         & 2,115        \\
                                   &                        & honggfuzz             & 7,253           & 2,273          & 0                & 52              & 7,253         & 2,325        \\
                                   &                        & libfuzzer             & 5,213           & 1,706          & 15               & 15              & 5,228         & 1,721        \\
                                   &                        & \g{\textbf{combined}} & \g{7,261}       & \g{2,330}      & \g{{ }}          & \g{{ }}         & \g{7,263}     & \g{2,363}    \\ \cmidrule{3-9}
\multirow{5}{*}{libevent}          & \multirow{5}{*}{17234} & afl                   & 423             & 117            & 0                & 7               & 423           & 124          \\
                                   &                        & aflpp                 & 427             & 119            & 0                & 5               & 427           & 124          \\
                                   &                        & honggfuzz             & 421             & 116            & 1                & 7               & 422           & 123          \\
                                   &                        & libfuzzer             & 418             & 122            & 0                & 3               & 418           & 125          \\
                                   &                        & \g{\textbf{combined}} & \g{427}         & \g{124}        & \g{{ }}          & \g{{ }}         & \g{427}       & \g{126}      \\ \cmidrule{3-9}
\multirow{5}{*}{re2}               & \multirow{5}{*}{21407} & afl                   & 11,492          & 5,193          & 18               & 115             & 11,510        & 5,308        \\
                                   &                        & aflpp                 & 11,485          & 5,209          & 68               & 183             & 11,553        & 5,392        \\
                                   &                        & honggfuzz             & 11,577          & 5,171          & 3                & 167             & 11,580        & 5,338        \\
                                   &                        & libfuzzer             & 11,003          & 5,009          & 4                & 81              & 11,007        & 5,090        \\
                                   &                        & \g{\textbf{combined}} & \g{11,586}      & \g{5,347}      & \g{{ }}          & \g{{ }}         & \g{11,586}    & \g{5,438}    \\ \cmidrule{3-9}
\multirow{5}{*}{woff2_new}         & \multirow{5}{*}{40914} & afl                   & 5,395           & 1,047          & 1                & 53              & 5,396         & 1,100        \\
                                   &                        & aflpp                 & 5,574           & 1,113          & 2                & 47              & 5,576         & 1,160        \\
                                   &                        & honggfuzz             & 5,555           & 1,082          & 5                & 75              & 5,560         & 1,157        \\
                                   &                        & libfuzzer             & 4,958           & 990            & 1                & 19              & 4,959         & 1,009        \\
                                   &                        & \g{\textbf{combined}} & \g{5,631}       & \g{1,193}      & \g{{ }}          & \g{{ }}         & \g{5,637}     & \g{1,247}    \\

    \end{tabular}
\endgroup

\label{table:big-table-asan}
\end{table*}

\begin{table}[tb]
\centering
\caption{
Mutants killed during 24 hour runs on 104 stubborn mutants for each subject (using ASAN).
}
\begingroup
    \fontsize{9pt}{9pt}\selectfont
    \begin{tabular}{ @{} l r r r r r @{} }
    Prog & Total & afl & aflpp & libfuzzer & honggfuzz \\ 
\midrule 
re2 & 104 & 0 & 0 & 0 & 0 \\ 
cares_parse_reply & 104 & 0 & 0 & 0 & 0 \\ 
woff2_new & 104 & 0 & 0 & 0 & 1 \\ 
curl & 104 & 0 & 0 & 1 & 0 \\ 
guetzli & 104 & 0 & 0 & 0 & 1 \\ 
libevent & 104 & 0 & 0 & 0 & 0 \\ 
cares_name & 66 & 0 & 0 & 0 & 0 \\ 

    \end{tabular}
\endgroup
\label{table:24-hour-runs}
\end{table}

\subsubsection{Manual Analysis of Mutations}
\label{sec:manual-analysis-mutations}
To assess whether mutations that we introduce result in a semantic change,
we manually examine 100 randomly selected mutations that were not found during the 24-hour runs.
Of these, we identify 11 mutations as equivalent,
5 as potentially leading to crashes, and the remaining 84 as introducing a semantic change, but unlikely to be detected by a simple crash oracle.

\subsection{RQ2: Evaluating the sanitizer contribution}
\begin{figure*}[t]
    \centering
    \includegraphics[width=\textwidth]{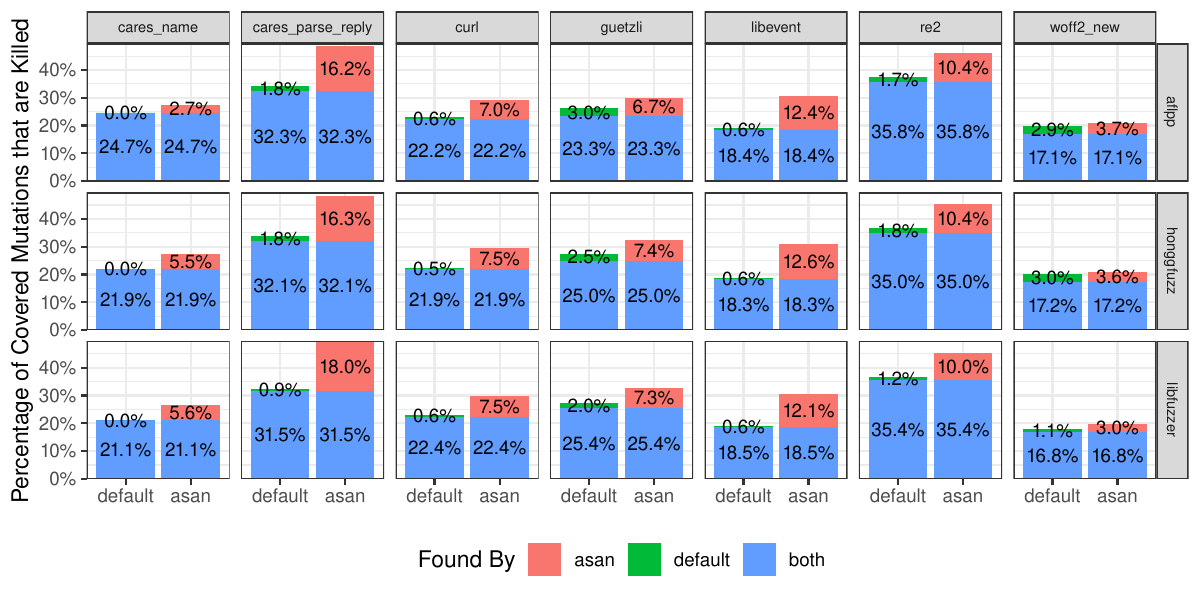}
    \caption{Percentage of covered mutants (in both experiments) that are killed with and without \ac{asan}. ``Both'' shows mutations killed with or without \ac{asan}, ``default'' those only found without \ac{asan}, and ``asan'' shows kills only found using \ac{asan}.
    }
    \label{fig:oracle-percentages-no-afl}
\end{figure*}

To answer this question, we re-run the previous experiment with \ac{asan}.
To analyze the results of this experiment, we compare the percentage of killed
mutants out of all covered mutations.
This is visualized for \aflpp, \honggfuzz and \libfuzzer in \Cref{fig:oracle-percentages-no-afl},
omitting \afl due to its inconsistencies between both oracles.
The full results can be seen in \Cref{table:big-table-asan}.
We see a clear increase of the number of killed mutants when using \ac{asan}.
Interestingly, enabling \ac{asan} also results in some crashes in the original
no longer being reported, which indicates a surprising effect of \ac{asan} instrumentation in some
cases.
Overall, with \ac{asan}, \aflpp found 9.1\% (+0.3\%) of mutations, \honggfuzz found 9.0\% (+0.3\%),
\afl found 8.5\% (+0.1\%), and \libfuzzer found 7.9\% (+0.4\%).
\begin{result}
\ac{asan} moderately increases the number of killed mutants.
\end{result}

\subsection{RQ3: Mutant and Vulnerability Coupling}
We analyzed the 100 most recent\footnote{as of September 5th, 2022} CVEs referencing GitHub commits that patch \verb|.c|, \verb|.cc|, \verb|.cpp| or \verb|.h| files.
A mutation is identified as \emph{coupled} to a vulnerability
if this mutation reintroduces the vulnerability into the patched program.
We will provide detailed results and justifications for this manual analysis along with our code.

For the evaluated CVEs, the way our mutations reintroduce bugs can be classified broadly as:
(1) The program behavior is modified such that the bug is reintroduced without side effects.
(2) The mutation reintroduces the bug, but also breaks some functionality.
(3) No mutation that reintroduces the bug can be found.

\begin{lstlisting}[float=h,belowskip=-1em,language=diff,caption={Patch for CVE-2022-34927.}, label={lst:patch-reversible}]
- if (instr[y].size < 29)
+ if (instr[y].size >= 4 && instr[y].size < 29)
\end{lstlisting}
An example patch for the first category can be seen in \Cref{lst:patch-reversible}.
The added check for \verb|size >=4| can be reversed by an \textsc{unsigned greater than equalto} mutation,
which will change the right-hand constant to \verb|0| here,
effectively re-introducing the original bug without side effects.

\begin{lstlisting}[float=h,belowskip=-1em,language=diff,caption={Patch for CVE-2022-39190.}, label={lst:patch-partly-reversible}]
+ if (nft_chain_is_bound(chain))
+     return -EINVAL;
\end{lstlisting}
A very common pattern among the studied patches is the introduction of a new branch checking an error condition.
These mostly fall into the second category because the buggy behavior can be triggered by inverting the branch condition (\textsc{redirect branch}), although this will result in the rejection of valid inputs.
\Cref{lst:patch-partly-reversible} shows an example for this situation:
After applying the mutation, only invalid \verb|chain|s are accepted, and valid input is discarded.

\begin{lstlisting}[float=h,belowskip=-1em,language=diff,caption={Patch for CVE-2022-29379.}, label={lst:patch-irreversible}]
- length = dir->length;
+ length += dir->length;
\end{lstlisting}
Patches in the third category sometimes require a more specialized mutation, such as the replacement of \verb|+=| with \verb|=| for the code in \Cref{lst:patch-irreversible}.
In other cases, patches may be impossible to revert due to information being lost (e.g., a deleted function call cannot be re-inserted).

For this analysis, we count vulnerabilities in categories (1) and (2) as coupled to a mutation, as per definition, the vulnerability is reintroduced.
As a result, 71 out of 100 analyzed CVEs are covered by our mutations.

\begin{result}
    The mutations induced by our mutation operators are~coupled to real faults.
\end{result}

\section{Discussion}
\label{sec:discussion}

Our paper demonstrates how to perform a principled, yet practical comparison of two fuzzers to determine whether
(1) using one is better than using the other or (2) using both combined is better than using
each in isolation (\Cref{fig:wayne}).
Furthermore, we can measure the improvement in a fuzzer compared to an older version
without biasing the results on previously discovered faults. %
Additionally, our approach can account for sanitizers and other strong oracles,
enabling a more principled comparison of oracles.
In summary, with this paper, we hope to encourage fuzzing researchers to develop better fuzzers without being unduly influenced by benchmarks.

\Mua is the gold standard of measuring test suite effectiveness in software
testing and comes with solid theory and empirical support.
The only restriction in using it for fuzzing so far has been the computational cost associated with it.
We have taken the first step in solving this issue, making \mua practical for fuzzing.
As proof that our approach indeed works in the real world,
we have compared four popular fuzzers on seven programs provided by
the OSS-Fuzz benchmark, the results of which we now present.

Our experiments largely confirm the fuzzer ranking established in previous
literature~\cite{hazimeh2020magma,metzman2021fuzzbench,zhang2022fixreverter}.
\aflpp leads the board, clearly outperforming \afl and \libfuzzer
(\honggfuzz was not part of the evaluation in~\cite{zhang2022fixreverter}).
Furthermore, our evaluation shows again that the fuzzers are very similar
and either \aflpp or \honggfuzz represent a solid choice.
As expected, the fuzzers in our evaluation have very limited targeted bug
finding capabilities, since there is nearly no improvement in the second phase.
The re-evaluation with \ac{asan} demonstrates the importance of improved oracles,
as the number of killed mutants increases.
It also shows that our approach is indeed capable of evaluating the complete tool chain.

\paragraph{Implications.}
Why should we trust \mua any more than the numerous metrics out
there, such as coverage measures~\cite{groce2014coverage},
defect-based benchmarks\cite{hazimeh2020magma,gavitt2016lava}, etc.?

As discussed in \Cref{sec:residual}, mutants that remain alive
represent possible undetected faults in the program, and with the detection of each
new mutant, that possibility decreases monotonically.
That is, unlike coverage, alive mutations are a good proxy measure for residual defects in a program.
Compared to defect-based benchmarks, mutation analysis minimizes bias and manual effort.
Finally, the mutations themselves can be studied to provide examples to improve fuzzers,
without the limitations of coverage or the bias of defect-based benchmarks.

Our results call for a reorientation of priorities in security testing. We have focused mostly on
improving coverage during fuzzing. However, it seems that there is a lot more to be gained by improving oracles instead.
This is not simple, however, and is known as the \emph{oracle problem} in software testing~\cite{barr2014oracle}.
The difficulty is that there is no general way to identify and extract the
\emph{intended logic} of a given program.
Some promising approaches exist that can help:
\begin{enumerate*}[label=(\arabic*), itemjoin={{; }}, itemjoin*={{; and }}, after={.}]
    \item metamorphic relations~\cite{segura2016survey,chen2016metamorphic} and equivalence modulo inputs~\cite{le2014compiler}
    \item differential oracles~\cite{mckeeman1998differential,gulzar2019perception}
    \item invariants such as those mined by Daikon~\cite{ernst2007daikon} and the like
\end{enumerate*}
We believe that future research should also be spent on improving the effectiveness of
these techniques rather than just pursuing the ever shrinking returns on improving coverage. 
\section{Limitations and Future Work}
\label{sec:limitations}

Our work is subject to the following important limitations.
\begin{description} %
    \item[Missing patterns and faults.] We rely on a set of mutation operators that we mined
    from real faults, which are unlikely to be exhaustive.
    This can be mitigated by future adaption of the supported mutations.
    \item[Cost of \mua.] While we reduced the cost of applying \mua to fuzzers,
    a further reduction would still improve the practicality of this approach.
    One such approach could involve sampling mutations, reducing computation time at the cost of accuracy.
    \item[Coupling effect.] We rely on the \emph{coupling effect} hypothesis in mutation
    analysis to ensure that the mutants we generate are similar to real faults.
    The coupling effect is well attested in literature~\cite{gopinath2017theory,just2014are}.
    However, relying on the coupling effect ignores subtle faults due to fault-interactions
    (between faults). While there is some evidence that such interactions
    are rare~\cite{kuhn2004software}, they are still important to address.
    One direction is exploring a larger neighborhood, with multiple mutations in a mutant.
    Unlike \supermutants, however, we need mutations
    that interact, and \emph{callability} that we used for \supermutants can provide a first level approximation.
    Indeed, given that \mua is a form of fuzz-testing the software test suite, the fuzzing
    community may be able to make better progress here.
    \item[Allocated Time for Phases.]
    Some fuzzers may not attempt to expand the coverage front and instead scan for possible vulnerability patterns, using static analysis.
    In such a case, the time provided for developing the coverage seed may not be useful to the fuzzer and may lead to unfair comparison.
    How to fairly compare such fuzzers without the overhead of full mutation analysis is an open question.
    (We note that while the coverage optimization may not work, the \supermutant optimization will work even in such cases.)
    A possible solution is to use the state-of-the-art coverage maximizing fuzzer on the original program and extract the coverage mutants from the given program.
    Next, use random sampling~\cite{gopinath2016limits} to identify a small set of coverage mutants, which when combined with the (complete set of) stubborn and other live mutants can serve to limit the overhead of full mutation analysis.
    Another option is to use the state-of-the-art coverage maximizing fuzzer and extract a mutant kill matrix.
    Then, identify the minimal test suite and the corresponding minimal mutant set, and use that minimal mutant set for evaluation of coverage mutants.
    We note that the relative merits of random sampling and minimal mutant set still needs to be calibrated~\cite{gopinath2016limits,gopinath2017mutation}.
    For random sampling, we note that a smaller sample size may suffice in theory~\cite{gopinath2015howhard}.
    However, the actual practical reduction in mutants is yet to be established (it is likely to be program-specific). 
   \item[Supermutants.] We use supermutants to check whether any fuzzer can find mutations that were
    not covered in the initial seed files.
    In doing this, we run the risk of fault masking.
    Hence, future work is to investigate the prevalence of fault masking and techniques to avoid fault masking.
    We, however, note that fault masking is likely to be rare.
    For one fault (say A) to mask the other (say B), the following conditions have to be fulfilled:
    (1) there should be no input that covers both in the coverage-maximizing phase,
    (2) fault A should never cause a crash independently (if it does, fault A would be removed and the remaining faults would be fuzzed separately),
    (3) there should be some inputs that cover both fault A and fault B, and on these inputs, fault A should induce just sufficient behavioral change such that, while the input may have caused a crash without fault A, the crash is removed when both are present.
    We believe that these constraints make fault masking due to supermutants rare.

    \item[Comparing Fuzzers without Sanitizers.] We find that sanitizers can significantly improve the effectiveness of fuzzers.
    However, \mua can allow us to go beyond merely accounting for the impact of sanitizers.
    We can compare the behavior (even coverage) of a mutant against the original using \emph{differential fuzzing} and identify whether a fuzzer was able to induce a change in behavior compared to the original even in the absence of a suitable sanitizer.
    This will allow us to develop effective fuzzers that focus on input generation independent of sanitizer development. Indeed,
    differential fuzzing by keeping track of coverage or behavior divergence between original and mutants can provide strong hints on which inputs hold promise in fuzzer guidance~\cite{groce2023first,vikram2023guiding}.
\end{description}

\section{Related Work}
\label{sec:related}
The comparison and evaluation of fuzzers is an important foundation for meaningfully improving fuzzers.
Several fuzzing platforms
exist~\cite{gavitt2016lava,hazimeh2020magma,metzman2021fuzzbench} that seek to provide a way to compare fuzzers under a common framework.

One such approach is LAVA~\cite{gavitt2016lava}, a tool to inject crash bugs
that can be triggered by finding specific values in unused parts of user-controlled input.
Later analysis showed that the introduced bugs are dissimilar to real-world vulnerabilities~\cite{klees2018evaluating},
are not coupled to real faults (reported CVEs)~\cite{bundt2021evaluating},
tend to overfit~\cite{zeller2019when}, and are ``solved''
by modern fuzzers~\cite{aschermann2019redqueen}.
In comparison, bugs from \mua are not guaranteed to be triggerable.
However, this is a trade-off
making it possible to create a comprehensive set of bugs, spanning from trivially detectable to subtle
and hard-to-detect ones.
A similar approach that can insert bugs into subjects is Evil Coder~\cite{pewny2016evilcoder}.
Potentially vulnerable source code locations are detected using data flow analysis,
focusing on user-controlled inputs that lead to sensitive functions.
A bug is introduced by removing security-relevant checks, such as input sanitization.
While our approach is not as targeted, mutation testing will not only generate similar bugs
but also a wider range of bugs.

The challenge binaries of the Cyber Grand Challenge (CGC)~\cite{darpa2016cgc}
are also sometimes used for comparison of fuzzers \cite{klees2018evaluating}.
The binaries were especially created for the CGC.
Hence, challenge binaries necessarily have a bias
to be used in the CGC and consist mostly of command line tools.

A recent benchmarking approach is Magma~\cite{hazimeh2020magma}.
It uses real-world vulnerabilities and re-inserts them into newer versions of the projects.
Additionally, it provides an assertion that tests if an input results in a state that would trigger the bug.
This assertion is used to measure bug detection capability.
As in other benchmarks, the number of bugs is limited because of the manual
effort to port them to the current version and has a necessary bias towards
bugs that can be re-inserted.

Another project is Fuzzbench~\cite{metzman2021fuzzbench}, using Google infrastructure.
Fuzzbench already provides coverage-based benchmarks and is working on supporting bug-based benchmarks~\cite{fuzzbench2023bugs}.

\textsc{FixReverter}~\cite{zhang2022fixreverter}
mines a restricted set of syntactic patterns from %
bug fixes that were associated with vulnerabilities and injects these bugs where the bug inducibility can be guaranteed, but has several limitations.
First, the researchers could identify only three general patterns accounting for 170 CVEs from a study of 814 CVEs (20.9\%).
Second, using patterns with semantic analysis to guarantee bug inducibility restricts the number of fixes that can be reverted. 
Such a guarantee also limits what kinds of bugs can be simulated, as the specific bug patterns and corresponding bug semantics that were mined represent only a small fraction of the possible bugs that can be present in a given program (in contrast to \mua). 
This might also open the door to fine-tuning fuzzers for specifically identifying such behavior. 
These drawbacks %
reduce the diversity of bugs and thus the effectiveness of the benchmark~\cite{gopinath2016limits,gopinath2017mutation}. 
Finally, \textsc{FixReverter} does not address the issue of fault-interactions and fault-masking.

B\"ohme et al.\@ suggest that coverage-based benchmarking can be
unreliable based on a comparison with curated bugs~\cite{bohme2022reliability}.
The paper illustrates the difficulty we face when we rely on an
external source of bugs.
In particular, because the bugs that the researchers rely on are external,
the distribution of such bugs is not related to the actual possibility of bugs
in the tested program.
To illustrate this, consider the following thought experiment:
Given a benchmark program that accepts inputs as JSON and a \blackbox random fuzzer.
The fuzzer will find lots of crashing bugs in the JSON parser itself but few in the program logic.
Any ranking using this source of bugs will favor fuzzers that find bugs in
the JSON parser when compared to, say, a grammar fuzzer that reaches the program internals.
Hence, the ranking based on finding such bugs is not a reliable indicator of fuzzer quality.
This is precisely what (unbiased) mutation analysis aims to correct.

\paragraph{Beyond Fuzzing.}
\Mua can also be applied for software verification tasks beyond fuzzing.
It can be used to evaluate the quality of static analysis tools~\cite{parveen2020mutation}, of type systems~\cite{gopinath2017good} (good type systems can make whole classes of errors non-representable), contracts~\cite{knuppel2021much}, and even the effectiveness of program proofs~\cite{jain2020mcoq}.

\section{Conclusion}
\label{sec:conclusion}

As fuzzing budget is limited, it is important to use fuzzers that
are better at finding faults.
The available benchmarks are, however, limited or biased
towards known bugs and are susceptible to overfitting and fine-tuning.

This paper demonstrates how the gold standard for measuring test suite quality---\mua---can be adapted
for fuzzing.
We show that two techniques, eliminating coverage mutants using static seed files and using
\supermutants for the remaining evaluation, can significantly reduce the computational expenditure necessary for \mua and can make mutation analysis feasible for fuzzing.
We investigated security faults and converted the identified patterns into security-specific
mutation operators, which were used for evaluation.
Using \mua, practitioners are no longer limited to specific curated benchmarks.
Instead, practitioners can evaluate fuzzers on the programs from a specific domain
before allocating resources for fuzzing.

Our evaluation demonstrates that with our technique, \mua can now be used for comparing
fuzzers in real-world programs.
Using \mua ensures that the practitioners can rely on the solid theory and decades
of empirical research, leading to better fuzzers and sanitizers.
The fine-grained results from mutation analysis can directly
help fuzzing practitioners to understand the deficiencies in current approaches and
take steps to correct them.
\todo{BM: Some information on the pattern table in the appendix where I am not sure how to integrate it understandably: The AtomicRMW pattern signal is only placed once, for the first such instruction found \textbf{(in the whole code)}. The mutation though is done on all atomic operators implemented and a signal is only added for the implemented operators. So it may happen that more signals are found in the mutated binary than in the original. It could also happen that there is an AtomicRMW instruction in the code but not with an implemented operator; then the original binary would have a found signal location, the mutated would not.
For the ATOMIC\_CMP\_XCHG pattern it is similar: only the first occurrence in a \textbf{function} is marked in the original binary, but all in the mutated binary.}

\section*{Acknowledgements}

This work was funded by the European Research Council (ERC) under the consolidator grant RS$^3$ (101045669) 
and the German Federal Ministry of Education and Research under the grant KMU-Fuzz (16KIS1523).
This work was partially supported with funds from the Bosch Research Foundation in the Stifterverband (Reference: T113/33825/19).
This work was supported by the Deutsche Forschungsgemeinschaft (DFG, German Research Foundation) under Germany’s Excellence Strategy – EXC-2092 CASA – 390781972.
We also thank the following colleagues and researchers for their help: Addison Crump, Joschua Schilling, Marcel Böhme, and Abhilash Gupta.

\renewcommand*{\bibfont}{\footnotesize}
\printbibliography

\appendix

\section{Appendix}

\subsection{Supplements for Evaluation}
Additional results as mentioned in the \Cref{sec:eval}.

\begin{table}[ht]
\centering
\caption{Number of mutants that were covered together with other mutants (i.e., mutants wrongly thought independent).}
\begingroup
    \fontsize{9pt}{9pt}\selectfont
    \begin{tabular}{@{} l r r r r @{}}
    Program           & afl   & aflpp & honggfuzz & libfuzzer \\ \midrule
cares\_name        & 4     & 0     & 0         & 0         \\
cares\_parse\_reply & 2     & 4     & 4         & 0         \\
curl              & 4,850 & 5,836 & 4,851     & 3,852     \\
guetzli           & 10    & 24    & 16        & 0         \\
libevent          & 0     & 2     & 0         & 0         \\
re2               & 39    & 66    & 37        & 47        \\
woff2\_new         & 26    & 46    & 56        & 48        \\

    \end{tabular}
\endgroup
\label{table:independent}
\end{table}

\paragraph{Supplementary Information for \Cref{table:independent}}
Any non-inde\-pendent mutants recorded during Phase~II are given in \Cref{table:independent}.
We find that except for \curl, the number of multi-covered mutants is minimal.
For \curl the interaction between mutations is caused by mutations that result in an error when handling http, these are then retried with the http2 protocol, covering functions and mutations there.
This kind of interactions can also occur in utility functions.

\begin{table*}
\centering
\caption{
    The number of mutants \textbf{Cov}ered or \textbf{Kill}ed for \textbf{Def}ault and \textbf{ASAN} experiments,
    for each mutation type and fuzzer. 
}
\begingroup
    \fontsize{7pt}{7pt}\selectfont
    
    \begin{tabular}{@{} p{2cm} l r r r r | p{2cm} l r r r r @{}}
        
        Mutation Type                                         & Fuzzer    & Cov Def & Cov Asan & Kill Def & Kill Asan & Mutation Type                                             & Fuzzer    & Cov Def & Cov Asan & Kill Def & Kill Asan \\ \midrule
\multirow{4}{\linewidth}{MALLOC}                      & afl       & 41      & 34       & 21       & 21        & \multirow{4}{\linewidth}{SIGNED LESS THAN}                & afl       & 447     & 543      & 104      & 95        \\
                                                      & aflpp     & 41      & 34       & 21       & 21        &                                                           & aflpp     & 670     & 556      & 116      & 103       \\
                                                      & honggfuzz & 41      & 34       & 20       & 21        &                                                           & honggfuzz & 672     & 555      & 111      & 100       \\
                                                      & libfuzzer & 40      & 34       & 14       & 21        &                                                           & libfuzzer & 571     & 471      & 100      & 91        \\ \midrule
\multirow{4}{\linewidth}{SIGNED GREATER THAN}         & afl       & 434     & 590      & 33       & 34        & \multirow{4}{\linewidth}{SIGNED LESS THAN EQUALTO}        & afl       & 2       & 2        & 0        & 0         \\
                                                      & aflpp     & 719     & 603      & 38       & 36        &                                                           & aflpp     & 2       & 2        & 0        & 0         \\
                                                      & honggfuzz & 725     & 604      & 42       & 35        &                                                           & honggfuzz & 2       & 2        & 0        & 0         \\
                                                      & libfuzzer & 614     & 526      & 28       & 29        &                                                           & libfuzzer & 2       & 2        & 0        & 0         \\ \midrule
\multirow{4}{\linewidth}{SIGNED GREATER THAN EQUALTO} & afl       & 4       & 4        & 0        & 0         & \multirow{4}{\linewidth}{FREE FUNCTION ARGUMENT}          & afl       & 923     & 887      & 918      & 879       \\
                                                      & aflpp     & 4       & 4        & 0        & 0         &                                                           & aflpp     & 981     & 945      & 976      & 939       \\
                                                      & honggfuzz & 4       & 4        & 0        & 0         &                                                           & honggfuzz & 944     & 909      & 940      & 905       \\
                                                      & libfuzzer & 3       & 3        & 0        & 0         &                                                           & libfuzzer & 876     & 843      & 870      & 837       \\ \midrule
\multirow{4}{\linewidth}{PTHREAD MUTEX}               & afl       & 3       & 3        & 0        & 0         & \multirow{4}{\linewidth}{SIGNED TO UNSIGNED}              & afl       & 825     & 1,172    & 32       & 30        \\
                                                      & aflpp     & 3       & 3        & 0        & 0         &                                                           & aflpp     & 1,393   & 1,197    & 38       & 35        \\
                                                      & honggfuzz & 3       & 3        & 0        & 0         &                                                           & honggfuzz & 1,407   & 1,209    & 35       & 34        \\
                                                      & libfuzzer & 3       & 2        & 0        & 0         &                                                           & libfuzzer & 1,201   & 1,060    & 32       & 29        \\ \midrule
\multirow{4}{\linewidth}{UNSIGNED TO SIGNED}          & afl       & 1,470   & 1,617    & 46       & 49        & \multirow{4}{\linewidth}{SWITCH SHIFT}                    & afl       & 568     & 692      & 6        & 9         \\
                                                      & aflpp     & 1,984   & 1,682    & 53       & 51        &                                                           & aflpp     & 903     & 785      & 5        & 13        \\
                                                      & honggfuzz & 1,981   & 1,666    & 52       & 55        &                                                           & honggfuzz & 825     & 711      & 2        & 9         \\
                                                      & libfuzzer & 1,736   & 1,489    & 40       & 48        &                                                           & libfuzzer & 718     & 635      & 3        & 9         \\ \midrule
\multirow{4}{\linewidth}{CALLOC}                      & afl       & 1       & 1        & 1        & 1         & \multirow{4}{\linewidth}{DELETE LOCAL STORE}              & afl       & 203     & 236      & 96       & 89        \\
                                                      & aflpp     & 1       & 1        & 1        & 1         &                                                           & aflpp     & 277     & 254      & 108      & 98        \\
                                                      & honggfuzz & 1       & 1        & 1        & 1         &                                                           & honggfuzz & 273     & 252      & 107      & 96        \\
                                                      & libfuzzer & 1       & 1        & 1        & 1         &                                                           & libfuzzer & 244     & 227      & 100      & 89        \\ \midrule
\multirow{4}{\linewidth}{UNSIGNED LESS THAN}          & afl       & 885     & 940      & 167      & 140       & \multirow{4}{\linewidth}{UNSIGNED GREATER THAN}           & afl       & 625     & 664      & 71       & 76        \\
                                                      & aflpp     & 1,121   & 970      & 182      & 145       &                                                           & aflpp     & 825     & 686      & 68       & 79        \\
                                                      & honggfuzz & 1,125   & 968      & 180      & 148       &                                                           & honggfuzz & 819     & 679      & 77       & 86        \\
                                                      & libfuzzer & 960     & 829      & 145      & 117       &                                                           & libfuzzer & 728     & 600      & 54       & 56        \\ \midrule
\multirow{4}{\linewidth}{UNSIGNED LESS THAN EQUALTO}  & afl       & 12      & 11       & 0        & 0         & \multirow{4}{\linewidth}{UNSIGNED GREATER THAN EQUALTO}   & afl       & 23      & 17       & 0        & 0         \\
                                                      & aflpp     & 15      & 13       & 0        & 0         &                                                           & aflpp     & 25      & 17       & 0        & 0         \\
                                                      & honggfuzz & 15      & 13       & 0        & 0         &                                                           & honggfuzz & 25      & 17       & 0        & 0         \\
                                                      & libfuzzer & 13      & 11       & 0        & 0         &                                                           & libfuzzer & 25      & 17       & 0        & 0         \\ \midrule
\multirow{4}{\linewidth}{COMPARE EQUAL TO}            & afl       & 2,367   & 2,643    & 1,055    & 1,164     & \multirow{4}{\linewidth}{SNPRINTF}                        & afl       & 14      & 9        & 0        & 0         \\
                                                      & aflpp     & 3,060   & 2,816    & 1,103    & 1,243     &                                                           & aflpp     & 14      & 9        & 0        & 0         \\
                                                      & honggfuzz & 2,992   & 2,741    & 1,074    & 1,209     &                                                           & honggfuzz & 14      & 9        & 0        & 0         \\
                                                      & libfuzzer & 2,620   & 2,390    & 974      & 1,088     &                                                           & libfuzzer & 13      & 9        & 0        & 0         \\ \midrule
\multirow{4}{\linewidth}{NEW ARRAY}                   & afl       & 2       & 21       & 2        & 13        & \multirow{4}{\linewidth}{SWITCH PLUS MINUS}               & afl       & 5,001   & 4,089    & 1,845    & 1,484     \\
                                                      & aflpp     & 21      & 21       & 2        & 13        &                                                           & aflpp     & 5,928   & 4,489    & 1,884    & 1,543     \\
                                                      & honggfuzz & 21      & 21       & 2        & 13        &                                                           & honggfuzz & 5,692   & 4,240    & 1,916    & 1,589     \\
                                                      & libfuzzer & 21      & 21       & 2        & 13        &                                                           & libfuzzer & 4,716   & 3,544    & 1,495    & 1,321     \\ \midrule
\multirow{4}{\linewidth}{REDIRECT BRANCH}             & afl       & 8,120   & 8,539    & 3,192    & 3,233     & \multirow{4}{\linewidth}{DELETE FUNCTION ARGUMENT}        & afl       & 466     & 433      & 465      & 432       \\
                                                      & aflpp     & 10,061  & 8,967    & 3,335    & 3,472     &                                                           & aflpp     & 469     & 436      & 468      & 435       \\
                                                      & honggfuzz & 9,963   & 8,806    & 3,311    & 3,432     &                                                           & honggfuzz & 469     & 436      & 468      & 435       \\
                                                      & libfuzzer & 8,534   & 7,578    & 2,819    & 3,038     &                                                           & libfuzzer & 447     & 415      & 445      & 414       \\ \midrule
\multirow{4}{\linewidth}{DELETE STORE PATTERN}        & afl       & 6,935   & 7,006    & 1,885    & 2,003     & \multirow{4}{\linewidth}{DELETE CALL INSTRUCTION PATTERN} & afl       & 1,545   & 1,607    & 184      & 741       \\
                                                      & aflpp     & 8,560   & 7,381    & 1,995    & 2,117     &                                                           & aflpp     & 2,122   & 1,757    & 193      & 838       \\
                                                      & honggfuzz & 8,471   & 7,256    & 2,002    & 2,088     &                                                           & honggfuzz & 2,022   & 1,666    & 192      & 800       \\
                                                      & libfuzzer & 7,060   & 6,253    & 1,752    & 1,881     &                                                           & libfuzzer & 1,699   & 1,436    & 166      & 708       \\ \midrule
\multirow{4}{\linewidth}{REASSIGN STORE INSTRUCTION}  & afl       & 2,421   & 2,241    & 368      & 372       &                                                           &           &         &          &          &           \\
                                                      & aflpp     & 2,878   & 2,354    & 400      & 388       &                                                           &           &         &          &          &           \\
                                                      & honggfuzz & 2,854   & 2,331    & 404      & 395       &                                                           &           &         &          &          &           \\
                                                      & libfuzzer & 2,300   & 1,940    & 354      & 345       &                                                           &           &         &          &          &           \\ \midrule

    \end{tabular}

\endgroup

\label{tbl:mut-type-fuzzer}

\end{table*}

\clearpage
\onecolumn

\subsection{Supplements for Approach}
The full list of mutation operators as mentioned in \Cref{sec:approach}.

\fontsize{6pt}{5.5pt}\selectfont
\def\arraystretch{1.5}
\begin{longtable}{@{} p{.18\textwidth} p{.38\textwidth} p{.38\textwidth} @{}}
\caption{List of all mutations used in our study.}\label{tbl:mutations} \\
Pattern Name & Description & Procedure \\ \midrule
\endfirsthead
\caption{(continued)} \\
Pattern Name & Description & Procedure \\ \midrule
\endhead
                         MALLOC &                                                                                                                                                                                      Mutating all malloc calls to achieve buffer overflow/out of bounds errors. &                                                                                                                                                                                                                                                                        We decrease allocated memory byte\_size in the malloc call by 16. \\
        FGETS MATCH BUFFER SIZE &                                                                                                                                                                                                     Mutating all fgets calls to achieve buffer overflow errors. &                                                                                                                                                                                                          We increase the size (n) parameter in the fgets call by increasing the value by 1 and then multiplying it by 5. E.g. 4->5->25. \\
               SIGNED LESS THAN &                                                                                                                Mutating all '<' comparisons either between two integer pointers or between 1 signed integer variable and an integer to achieve overflow errors. &                                                                                                                For pointer comparison, 8*4=32 is added to the right hand side pointer in the comparison. For integer comparison, the integer on the right hand side is squared if larger than 1024 or smaller than 2, else 32 is added. \\
            SIGNED GREATER THAN &                                                                                                               Mutating all '>' comparisons either between two integer pointers or between 1 signed integer variable and an integer to achieve underflow errors. &                                                For pointer comparison, 8*4=32 is subtracted from the right hand side pointer in the comparison. For integer comparison, either the sqrt is taken for integers > 1024*1024, halved for integers > 1024 and either 0 is returned or 32 is substracted, whatever gives the largest result. \\
       SIGNED LESS THAN EQUALTO &                                                                                                               Mutating all '<=' comparisons either between two integer pointers or between 1 signed integer variable and an integer to achieve overflow errors. &                                                                                                                For pointer comparison, 8*4=32 is added to the right hand side pointer in the comparison. For integer comparison, the integer on the right hand side is squared if larger than 1024 or smaller than 2, else 32 is added. \\
    SIGNED GREATER THAN EQUALTO &                                                                                                              Mutating all '>=' comparisons either between two integer pointers or between 1 signed integer variable and an integer to achieve underflow errors. &                                                For pointer comparison, 8*4=32 is subtracted from the right hand side pointer in the comparison. For integer comparison, either the sqrt is taken for integers > 1024*1024, halved for integers > 1024 and either 0 is returned or 32 is substracted, whatever gives the largest result. \\
         FREE FUNCTION ARGUMENT &                                                                                                                          Mutating all functions that receive a pointer type function argument to achieve double free and possibly illegal memory access errors. &                                                                                                                                                                                             We check for functions that receive a pointer type argument. Before returning at the end of the function, one argument per mutant is freed. \\
                  PTHREAD MUTEX &                                                                                                                                                                                Mutating all pthread\_lock and pthread\_unlock calls to achieve data races errors. &                                                                                                                                                                                                                                                           We remove all pthread\_lock and pthread\_unlock calls in a function per mutant. \\
                ATOMIC CMP XCHG &                                                                                                                                                                                                    Mutating all atomic compare exchanges to achieve data races. &                                                                                                                                                               If we have at least one atomic cmpxchg instruction, we replace all atomic cmpxchg return success values (the element with index 1 in the result array) by 1 per function. \\
              ATOMICRMW REPLACE &                                                                                                                                                                                                      Mutating all atomicrmw instructions to achieve data races. &                                                                                                               Takes the given atomic instruction and replaces it with its non-atomic counterpart for the following instructions: ADD, SUB, AND, OR, XOR, FADD, FSUB. For other operators no mutation is done, the mutant is equivalent. \\
             SIGNED TO UNSIGNED &                                                                                                                                                                            Mutating all signed integer comparisons to achieve overflow and out of bound errors. &                                                                                                                               Each of the four integer comparison predicates - ICMP\_SGT, ICMP\_SGE, ICMP\_SLT, ICMP\_SLE are transformed into the corresponding unsigned predicates - ICMP\_UGT, ICMP\_UGE, ICMP\_ULT, ICMP\_ULE respectively. \\
             UNSIGNED TO SIGNED &                                                                                                                                                                         Mutating all unsigned integer comparisons to achieve overflow and out of bounds errors. &                                                                                                                               Each of the four integer comparison predicates - ICMP\_UGT, ICMP\_UGE, ICMP\_ULT, ICMP\_ULE are transformed into the corresponding unsigned predicates - ICMP\_SGT, ICMP\_SGE, ICMP\_SLT, ICMP\_SLE respectively. \\
                   SWITCH SHIFT &                                                                                                                                                                                          Mutating all shift calls to achieve overflow and out of bounds errors. &                                                                                                                                                                                                                                                                       Replaces an arithmetic shift with a logical shift and vice versa. \\
                         CALLOC &                                                                                                                                                                                         Mutating all calloc calls to achieve overflow and out of bounds errors. &                                                                                                                                                                                                                                                                                          The size parameter's value is decreased by 16. \\
             DELETE LOCAL STORE &                                                                                                                                                                     Mutating all stores on a local variable in one function to achieve uninitialization errors. &                                                                                                                                                                                                                                                                                                              The store call is removed. \\
             UNSIGNED LESS THAN &                                                                                                              Mutating all '<' comparisons either between two integer pointers or between 1 unsigned integer variable and an integer to achieve overflow errors. &                                                                                                                For pointer comparison, 8*4=32 is added to the right hand side pointer in the comparison. For integer comparison, the integer on the right hand side is squared if larger than 1024 or smaller than 2, else 32 is added. \\
          UNSIGNED GREATER THAN &                                                                                                             Mutating all '>' comparisons either between two integer pointers or between 1 unsigned integer variable and an integer to achieve underflow errors. &                                                For pointer comparison, 8*4=32 is subtracted from the right hand side pointer in the comparison. For integer comparison, either the sqrt is taken for integers > 1024*1024, halved for integers > 1024 and either 0 is returned or 32 is substracted, whatever gives the largest result. \\
     UNSIGNED LESS THAN EQUALTO &                                                                                                             Mutating all '<=' comparisons either between two integer pointers or between 1 unsigned integer variable and an integer to achieve overflow errors. &                                                                                                                For pointer comparison, 8*4=32 is added to the right hand side pointer in the comparison. For integer comparison, the integer on the right hand side is squared if larger than 1024 or smaller than 2, else 32 is added. \\
  UNSIGNED GREATER THAN EQUALTO &                                                                                                            Mutating all '>=' comparisons either between two integer pointers or between 1 unsigned integer variable and an integer to achieve underflow errors. &                                                For pointer comparison, 8*4=32 is subtracted from the right hand side pointer in the comparison. For integer comparison, either the sqrt is taken for integers > 1024*1024, halved for integers > 1024 and either 0 is returned or 32 is substracted, whatever gives the largest result. \\
    INET ADDR FAIL WITHOUTCHECK &                                                                                                                                                       Mutating all calls to the libc function inet\_addr to achieve unhandled non-established connection errors. & Replaces all uses of the function return value to the failure value. Also removes the function call from the corpus as a fail of the function call should be simulated. Furthermore, the comparison instructions are flipped, s.t. on failure the 'correct' path is taken, i.e. we simulate a missing check for the error return value. \\
               COMPARE EQUAL TO &                                                                                                                                                                                                      Mutating all '==' comparisons between two integers to '='. &                                                                                                                                           The value of integer on the right hand side is assigned to the variable on the left. The condition passes and the inside block is executed as long as the value on the RHS is not equal to 0. \\
                         PRINTF &                                                                                                                                                                       Mutating printf such that the format string gets already filled and then plainly printed. &                                            Mutating printf such that the format string is already filled on printing, so instead of calling printf('\%d \%s', 10, string); we simulate the call printf('10 <string-value>');. This might cause illegal memory accesses and printing of secrets if the string argument is user controlled. \\
                        SPRINTF &                                                                                                                                                                      Mutating sprintf such that the format string gets already filled and then plainly printed. &                         Mutating sprintf such that the format string is already filled on printing, so instead of calling sprintf('\%d \%s', buffer, 10, string); we simulate the call sprintf('10 <string-value>', buffer);. This might cause illegal memory accesses and printing of secrets if the string argument is user controlled. \\
                       SNPRINTF &                                                                                                                                                                     Mutating snprintf such that the format string gets already filled and then plainly printed. &         Mutating snprintf such that the format string is already filled on printing, so instead of calling snprintf('\%d \%s', size, buffer, 10, string); we simulate the call snprintf('10 <string-value>',  size, buffer);. This might cause illegal memory accesses and printing of secrets if the string argument is user controlled. \\
                      NEW ARRAY &                                                                                                                                                                               Mutating new[] in (only) cpp files such that the array is allocated lesser memory &                                                                                                                                                                                                                                                                         We decrease allocated memory size in the 'new' call by 5 units. \\
              SWITCH PLUS MINUS &                                                                                                                                                                                                       Changing a '+' operator to a '-' operator and vice versa. &                                                                                                                                                                                                                                           Changing a '+' operator to  a '-' operator regardless for integer and floating point numbers. \\
                REDIRECT BRANCH &                                                                                                                                                                                                  Negate the result of the branching condition before branching. &                                                                                                                                                                                                                                                  Redirecting the control flow by negating the result of the condition before branching. \\
       DELETE FUNCTION ARGUMENT & Mutating all functions in (only) cpp files that receive a pointer type function argument to achieve double delete and possibly illegal memory access errors. N.B. - Can possibly lead to a memory leak when delete is called for arrays instantiated with new[] &                                                                                                                                                                                          We check for functions that receive a pointer type argument. Before returning at the end of the function, one argument per mutant is deleted.  \\
           DELETE STORE PATTERN &                                                                                                                                                                          Deletes all store instructions one by one to simulate a forgotten variable assignment. &                                                                                                                                                                                                                 Find a store instruction and delete it. As there are no further dependencies on the store, there is nothing else to do. \\
DELETE CALL INSTRUCTION PATTERN &                                                                                                                                            Deletes all call instructions without return value assignment one by one to simulate a forgotten call to a function. &                                                                                                                                                                       Find a call instruction without return value assignment and delete it. As there are no further dependencies on the call instruction, there is nothing else to do. \\
     REASSIGN STORE INSTRUCTION &                                                                                                                                                                                       Reassigns the value of a previous store with the same type in this store. &                                                                                                                      Checks if in this basic block is another store with the same types used and assigns the first operand of the previous store to the memory location denoted by the second operand of the store we are currently at. \\
\end{longtable}

\twocolumn

\end{document}